\documentclass[11pt]{article}%

\usepackage{arxiv}

\usepackage[utf8]{inputenc} %
\usepackage[T1]{fontenc}    %
\usepackage[hidelinks]{hyperref}       %
\usepackage{url}            %
\usepackage{booktabs}       %
\usepackage{amsmath}
\usepackage{amsthm}
\usepackage{amssymb}
\usepackage{amsfonts}       %
\usepackage{nicefrac}       %
\usepackage{microtype}      %
\usepackage{graphicx}
\usepackage[round]{natbib}
\usepackage{doi}

\usepackage{numprint}%
\npdecimalsign{.}
\npthousandsep{,}

\usepackage{dsfont} %

\usepackage{color} %

\graphicspath{{figures/}}

\theoremstyle{plain}

\newtheorem{theorem}{Theorem}[section]

\theoremstyle{remark}
\newtheorem{definition}[theorem]{Definition}

\newcommand{\Reals}{\mathbb{R}}

\newcommand{\Prob}{\mathbb{P}}
\newcommand{\E}{\mathbb{E}}
\newcommand{\1}{\mathds{1}}

\DeclareMathOperator*{\pa}{pa}
\DeclareMathOperator*{\An}{An}
\DeclareMathOperator*{\an}{an}
\DeclareMathOperator*{\RV}{RV}

\title{Causal Modelling of Heavy-Tailed Variables and Confounders with Application to River Flow}

\date{} 					%

\author{Olivier~C.~Pasche\\
    Research Center for Statistics, University of Geneva, Switzerland,\\
    Institute of Mathematics, EPFL, 1015 Lausanne, Switzerland,\\
    \href{mailto:olivier.pasche@unige.ch}{\texttt{olivier.pasche@unige.ch}}\\
    \And
    Val\'erie~Chavez-Demoulin\\
    Faculty of Business and Economics, University of Lausanne, Switzerland,\\
    \href{mailto:valerie.chavez@unil.ch}{\texttt{valerie.chavez@unil.ch}}\\
    \And
    Anthony~C.~Davison\\
    Institute of Mathematics, EPFL, 1015 Lausanne, Switzerland,\\
    \href{mailto:anthony.davison@epfl.ch}{\texttt{anthony.davison@epfl.ch}}\\
}

\renewcommand{\headeright}{}
\renewcommand{\undertitle}{}%
\renewcommand{\shorttitle}{Causal Modelling of Heavy-Tailed Variables and Confounders}

\hypersetup{
pdftitle={Causal Modelling of Heavy-Tailed Variables and Confounders with Application to River Flow},
pdfsubject={stat.ME},
pdfauthor={Olivier C. Pasche, Val\'erie Chavez-Demoulin, Anthony C. Davison},
pdfkeywords={Causation, Causal tail coefficient, Confounder, Extreme value statistics, Generalized Pareto distribution},
}

\begin{document}
\maketitle

\begin{abstract}
Confounding variables are a recurrent challenge for causal discovery and inference. In many situations, complex causal mechanisms only manifest themselves in extreme events, or take simpler forms in the extremes.  Stimulated by data on extreme river flows and precipitation, we introduce a new causal discovery methodology for heavy-tailed variables that allows the effect of a known potential confounder to be almost entirely removed when the variables have comparable tails, and also decreases it sufficiently to enable correct causal inference when the confounder has a heavier tail.  We also introduce a new parametric estimator for the existing causal tail coefficient and a permutation test. Simulations show that the methods work well and the ideas are applied to the motivating dataset.
\end{abstract}


\section{Introduction}
The field of causal inference has developed massively in recent decades \citep[e.g.,][]{pearl09,peters.etal:2017}, with much recent work on the detection of causality from observational data \citep[e.g.,][]{maathuis16}.
Most of this literature concerns central quantities such as expectations, but certain causal mechanisms manifest themselves only in rare events and/or may simplify in distribution tails.  Standard methods of causal inference are ill-suited for such situations, and recent work has begun to link causality and  extreme value theory. Examples are \citet{Gissibl.Kluppelberg:2018}, who  define recursive max-linear models on directed acyclic graphs, \cite{Kluppelberg21}, who propose a scaling technique to determine the causal order of the variables in such graphs, \citet{kiriliouk}, who use multivariate generalized Pareto distributions to study probabilities of necessary and sufficient causation as defined in the counterfactual theory of Pearl, and \citet{mhalla2018}, who construct a causal inference method for tail quantities relying on Kolmogorov complexity of extreme conditional quantiles.  See surveys by  \cite{naveau2020} on extreme event attribution and by \cite{Engelke2020} on the detection and modeling of sparse patterns in extremes. 

Our work stems from that of \citet{Gnecco2019}, who propose an estimator of the causal tail coefficient and an algorithm that, under mild conditions,  consistently retrieves a causal order on an underlying graph even in the presence of hidden confounders. Such an order helps to exclude some causal structures, but does not provide evidence for the existence of a specific structure, as in general a given order is causal for several possible graphs; in particular, all orders are causal for the empty graph corresponding to absence of causality.
Although it is asymptotically invariant to hidden confounders,  this  estimator can suffer from confounding in finite samples when inference on the direct relationship between two variables is needed, when these effects are too strong or when the confounders have heavier tails than the two variables.

This paper addresses a central challenge in causal inference: the presence of confounders. In theoretical development it is often assumed that all the relevant variables are observed and can be included in the model, but in practice one can rarely be sure of this. The available variables are often subject to external influences, observed or unobserved, that affect the variables of interest and can make it harder or even impossible to infer a correct causal relationship.  Our goals are to mitigate the effect of a set of known confounders on an extremal causal analysis by treating them as covariates, and to present a permutation test for direct causality between the two observed variables. Our approach relaxes the assumption of \citet{Gnecco2019} that the confounders have the same tail index as the two main variables of interest, and thus encompasses a much broader range of situations, such as that in our application. Such a model enables causal discovery and inference for a greater variety of situations.

\begin{figure}[!t]
\centering
\includegraphics[width=\textwidth]{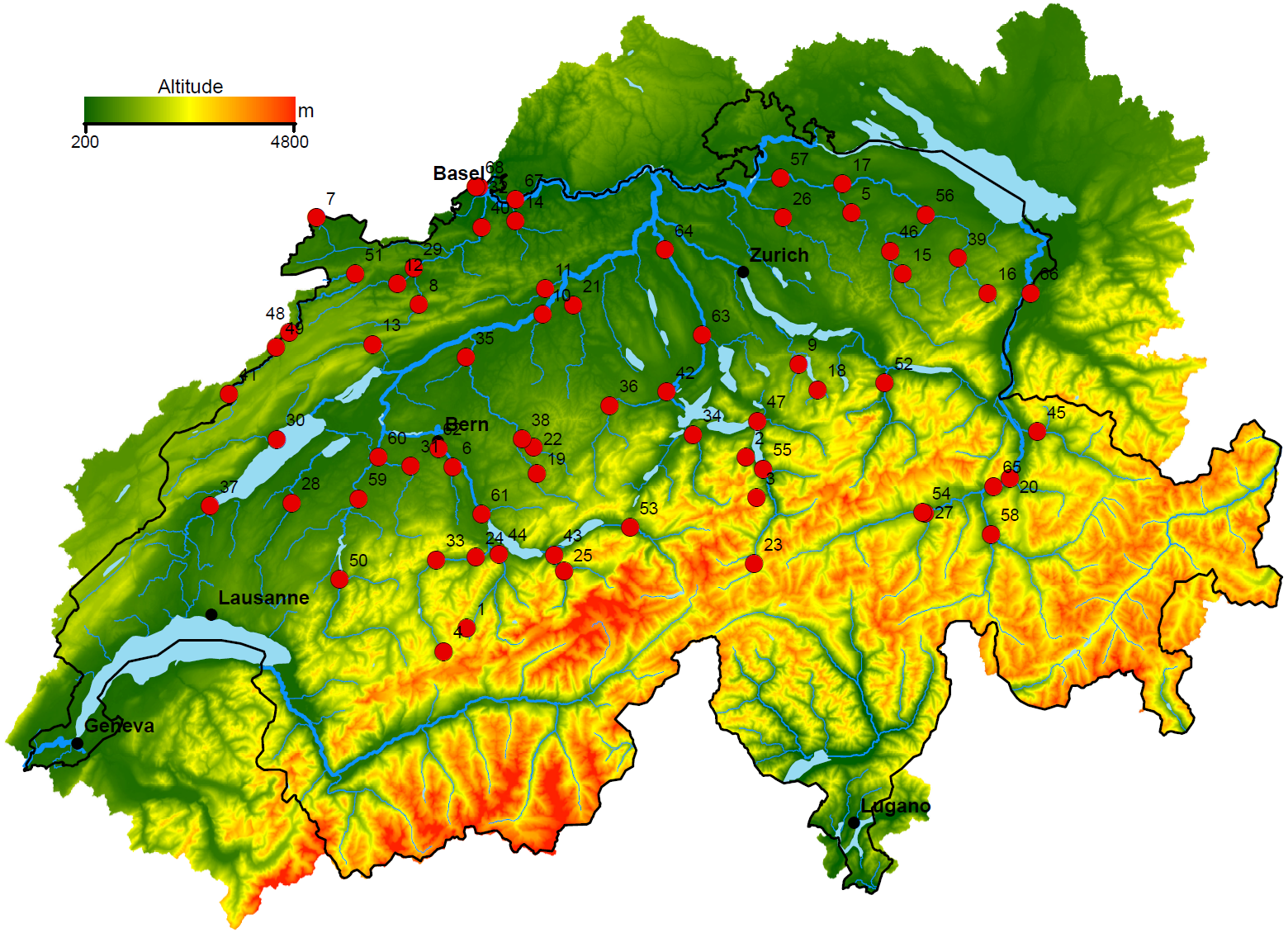}
\caption{Topographic map of Switzerland showing the $68$ gauging stations (red dots) along the Rhine, the Aare and their tributaries. Water flows towards station $68$. Adapted from~\citet{Engelke2020}.}
\label{fd:chstations}
\end{figure}

Our work was stimulated by average daily discharge data from 68 gauging stations along the Rhine and Aare catchments in Switzerland, see Figure~\ref{fd:chstations}. The data were collected by the Swiss Federal Office for the Environment (\href{https://hydrodaten.admin.ch}{hydrodaten.admin.ch}), but were provided by the authors of~\citet{Engelke2020}, with some useful preliminary insights. We focus on the causal relationship between extreme discharges, for which precipitation is an obvious confounder, and use daily precipitation data from $105$ meteorological stations, provided by the Swiss Federal Office of Meteorology and Climatology, MeteoSwiss (\href{https://gate.meteoswiss.ch/idaweb}{gate.meteoswiss.ch/idaweb}). Unlike in our simulation experiments, we know neither the true tail properties  of the discharges and precipitation nor the effect of the confounder.
We use precipitation as a covariate in our test, allowing inference on the direct causal relationships between discharges for the majority of the station pairs, with at least $95\%$ estimated confidence, which was impossible without our proposed approach.

The paper is organised as follows. Section~\ref{s:ctc} discusses the causal tail coefficient, its interpretation and its properties. Section~\ref{s:pctc} introduces a new parametric estimator for it based on generalized Pareto modelling of threshold excesses, which allows a known confounder to be used as a covariate. A simulation study in Section~\ref{s:simul} underlines the strengths and limitations of the two estimators. Section~\ref{s:test} presents a  permutation test intended to detect direct causality between two heavy-tailed variables, which is also assessed via simulation.  Section~\ref{s:rivers} applies the methodology to the river discharges, and Section~\ref{s:conclu} gives a brief discussion.

\newpage

\section{Causal Tail Coefficient and its Estimation}\label{s:ctc}

\subsection{Existing Work}

We first give some basic notions needed to describe the setting in which causal relationships between random variables can be recovered. 

\begin{definition}\label{df:scm}
A \emph{linear structural causal model (LSCM)} over a set of random variables $X_1,\dots,X_p$ satisfies
\begin{equation*}
    X_j=\sum_{k\in \pa(j)}\beta_{jk}X_k+\varepsilon_j, \quad j\in V,%
\end{equation*}
where $V:=\{1,\dots,p\}$ is a set of nodes representing the corresponding random variables, $\pa(j)\subseteq V$ is the set of parents of $j$, $\beta_{jk}\in\Reals\setminus\{0\}$ is called the \emph{causal weight} of node $k$ on node $j$, and $\varepsilon_1,\dots,\varepsilon_p$ are jointly independent noise variables. We suppose that the \emph{associated graph} $G=(V,E)$, in which the directed edge $(i,j)\in V\times V$ belongs to $E$ if and only if $i\in\pa(j)$, is a directed acyclic graph (DAG).
\end{definition}

In a DAG $G=(V,E)$, we say that $i\in V$ \emph{is an ancestor of} $j\in V$ in $G$, if there exists a directed path from $i$ to $j$. The set of the ancestors of $j$ in $G$ is denoted by $\An(j,G)$, and we define $\an(j,G):=\An(j,G)\setminus\{j\}$. In a LSCM over random variables $X_1,\dots,X_p$, with associated DAG $G=(V,E)$, we say that $X_i$ \emph{causes} $X_j$, if $i\in\an(j,G)$. We call $X_i$ a \emph{confounder} (or \emph{common cause}) of $X_j$ and $X_k$ if there exist directed paths from $i$ to $j$ and from $i$ to $k$ in $G$ that do not include $k$ and $j$, respectively. We say that there is \emph{no causal link} between $X_i$ and $X_j$ if $\An(i,G)\cap\An(j,G)=\emptyset$.  For any $i,j\in V$ we let $\beta_{i\rightarrow j}$ denote the sum of the products of the causal weights along the distinct directed paths from vertex $i$ to vertex $j$; we set $\beta_{j\rightarrow j}:=1$ and $\beta_{i\rightarrow j}:=0$ if $i\notin\An(j,G)$.

Let $X_i$ and $X_j$ be random variables from a LSCM with respective distributions $F_i$ and $F_j$. The \emph{causal (upper) tail coefficient}\ of a random variable $X_i$ on another random variable $X_j$ is defined as \citep{Gnecco2019} 
\begin{equation}\label{eq:ctc}
    \Gamma_{ij}:=\lim_{u\to 1^-}\E\left\{F_j(X_j)\mid F_i(X_i)>u\right\},
\end{equation}
if the limit exists. This coefficient lies between zero and one and captures the causal influence of $X_i$ on $X_j$ in their upper tails: if $X_i$ has a linear causal effect on $X_j$, $\Gamma_{1,2}$ will be close to unity. The coefficient is asymmetric, as extremes of $X_j$ need not lead to extremes of $X_i$, and in that case, $\Gamma_{ji}$ will be appreciably smaller than $\Gamma_{ij}$. As $\Gamma_{ij}$ only depends on the rescaled margins of the variables, it is invariant to monotone increasing marginal transformations.

If both tails are of interest, the causal tail coefficient can be generalized to capture the causal effects in both directions, by considering the \emph{symmetric causal tail coefficient}\ of $X_i$ on $X_j$, i.e., 
\begin{equation*}
    \Psi_{ij}:=\lim_{u\to 1^-}\E\left[\rho\left\{F_j(X_j)\right\}\mid\rho\left\{F_i(X_i)\right\}>u\right]
\end{equation*}
if the limit exists, where $\rho : x\mapsto \lvert 2x-1\rvert$.
As $F_i(X_i)\sim {\rm Unif}(0,1)$,
\begin{equation*}
    \Psi_{ij}=\underbrace{\lim_{u\to 1^-}\dfrac{1}{2}\E\left[\rho\left\{F_j(X_j)\right\}\mid F_i(X_i)>u\right]}_{=:\Psi_{ij}^+} + \underbrace{\lim_{u\to 0^+}\dfrac{1}{2}\E\left[\rho\left\{F_j(X_j)\right\}\mid F_i(X_i)<u\right]}_{=:\Psi_{ij}^-}.
\end{equation*}
The interpretation and properties of $\Psi_{ij}$ are similar to those of $\Gamma_{ij}$. The symmetric version captures the causal influence of $X_i$ on $X_j$ in both of their tails. 

For simplicity we focus on $\Gamma_{ij}$ in this paper, though all of our results and methods can be generalized to both tails by considering $\Psi_{ij}$ instead, if the assumptions for the upper tails are also satisfied in the lower tails of the variables considered.

Before stating the theorem that describes how the underlying causal relationships in a set of random variables can be recovered, we define the concept of regular variation.

\begin{definition}\label{df:rvf}
A positive measurable function $f$ is said to be \emph{regularly varying} with index $\alpha\in\Reals$, written $f\in\RV_\alpha$, if for all $c>0$, $\lim_{x\to\infty}f(cx)/f(x)=c^\alpha$.
If $f\in\RV_0$, then $f$ is said to be \emph{slowly varying}.
\end{definition}

\begin{definition}\label{df:rvx}
The random variable $X_j$ is said to be \emph{regularly varying} with index $\alpha>0$, if, for some $\ell\in\RV_0$, $\Prob(X_j>x)\sim\ell(x)x^{-\alpha}$ as $x\to\infty$.
\end{definition}
Independent regularly varying random variables $X_1,\dots,X_p$ are said to have \emph{comparable upper tails} if there exist $c_1,\dots,c_p>0$, $\alpha>0$ and $\ell\in\RV_0$ such that, for each $j\in\{1,\dots,p\}$, $\Prob(X_j>x)\sim c_j\ell(x)x^{-\alpha}$ as $x\to\infty$.

The following theorem describes how the causal relationships underlying  a set of random variables can be recovered from their causal tail coefficients.
\begin{theorem}[\citeauthor{Gnecco2019}, \citeyear{Gnecco2019}]\label{thm:ctc1val}
Let $X_1,\dots,X_p$ be random variables from a LSCM, with associated directed acyclic graph $G=(V,E)$ and suppose that%
\begin{enumerate}
\item[(a)] the coefficients $\beta_{jk}$ of the linear structural causal relationship $X_j=\sum_{k\in \pa(j,G)}\beta_{jk}X_k+\varepsilon_j$ are strictly positive for all $j\in V$ and $k\in\pa(j,G)$, and %
\item[(b)] the real-valued noise variables $\varepsilon_1,\dots,\varepsilon_p$ are independent and regularly varying with comparable upper tails.
\end{enumerate}
Then the values of $\Gamma_{ij}$ and $\Gamma_{ji}$ allow one to distinguish between the different possible causal relationships between $X_i$ and $X_j$ summarized in Table~\ref{t:ctc1}.

\begin{table}[!ht]
    \caption{Equivalence of the possible values of $\Gamma_{ij}$ and $\Gamma_{ji}$ with the underlying causal relationship between $X_i$ and $X_j$.}%
    \label{t:ctc1}
    \centering
    \begin{tabular}{@{}|l|ccc|@{}}%
         \hline
         & $\Gamma_{ji}=1$ & $\Gamma_{ji}\in(1/2,1)$ & $\Gamma_{ji}=1/2$ \\
         \hline
         $\Gamma_{ij}=1$ & & $X_i$ causes $X_j$ & \\
         $\Gamma_{ij}\in(1/2,1)$ & $X_j$ causes $X_i$ & common cause only & \\
         $\Gamma_{ij}=1/2$ & & & no causal link \\
         \hline
    \end{tabular}
\end{table}
\end{theorem}

Under the theorem's assumptions, the blank entries in Table~\ref{t:ctc1} cannot occur. Theorem~\ref{thm:ctc1val} is generalizable to the $\Psi_{ij}$ variant of the coefficient and possibly negative $\beta_{ij}$ values if the assumptions are also satisfied in the lower tails of the variables.

\citet{Gnecco2019} show that under the setup and assumptions of Theorem~\ref{thm:ctc1val}, the causal tail coefficient~\eqref{eq:ctc} for any distinct $i,j\in V$, and with $A_{ij}:=\An(i,G)\cap\An(j,G)$, is
\begin{equation}\label{eq:lemctc1}
    \Gamma_{ij}=\dfrac{1}{2}+\dfrac{1}{2}\dfrac{\sum_{h\in A_{ij}}\beta_{h\rightarrow i}^\alpha}{\sum_{h\in\An(i,G)}\beta_{h\rightarrow i}^\alpha}.
\end{equation}
Without loss of generality we set $i=1$ and $j=2$ in what follows, and thus consider the causal effect of $X_1$ on $X_2$.

If  $\left\{(X_{i,1},X_{i,2})\right\}_{i=1}^n$ are independent replicates of $(X_1,X_2)$, with the random variables $X_i$ and $X_j$ from the LSCM, then the \emph{non-parametric estimator} of $\Gamma_{1,2}$ is defined to be
\begin{equation}\label{eq:npctc1}
    \hat{\Gamma}_{1,2}=\dfrac{1}{k}\sum_{i=1}^n\hat{F}_2(X_{i,2})\1(X_{i,1}>X_{(n-k),1}) 
\end{equation}
for some $k\in\{1,\ldots, n-1\}$, where $\1(\cdot)$ denotes the indicator function, $X_{(h),1}$ denotes the $h^\text{th}$ order statistic and $\hat{F_j}$ is the empirical cumulative distribution function of $X_j$, i.e., 
\begin{equation*}
    \hat{F}_j(x)=\dfrac{1}{n}\sum_{i=1}^n\1(X_{i,j}\leq x),\quad j=1,2.
\end{equation*}
This estimator is the empirical counterpart to~\eqref{eq:ctc}, as $X_{(h),1}=\hat{F}_1^\leftarrow(h/n)$ is a quantile of the corresponding empirical distribution. The value of $k$ controls the number of data pairs in the upper tail of $X_1$ that contribute to the estimator. Under the assumptions of Theorem~\ref{thm:ctc1val} and a ``very mild assumption that is satisfied by most univariate regularly varying distributions of interest'', estimator~\eqref{eq:npctc1} is consistent as $n\to\infty$, for a choice of $k$ such that $k\to\infty$ and $k/n\to 0$ \citep{Gnecco2019}.

\subsection{Practical Limitations}

A strength of the causal tail coefficient approach is its asymptotic robustness to hidden confounders. Studies of causation frequently presuppose that all the relevant variables have been observed, which is usually moot, but Theorem~\ref{thm:ctc1val} holds even when some variables in the underlying LSCM are unobserved. This capacity to deal with confounders both when studying the causal relationship between two variables and when retrieving a causal order is not generally shared by other approaches in causal inference, as argued by~\citet[Section~4.2]{Gnecco2019}, but the unobserved variables must satisfy a regular variation assumption that is hard to check and may be unrealistic. In practice, moreover, the tail behaviour of the confounders may differ from that of $X_1$ and $X_2$, violating assumption (b) of Theorem \ref{thm:ctc1val}. In our motivating setting, for example, the tail of the confounder, precipitation, may not behave like the tails of the river discharges.  This problem worsens when the confounder has a heavier tail than the variable of interest. Furthermore, distinguishing between different causal situations using empirical estimates may be difficult; an increase in the strength of the causal effect of a common confounder of $X_1$ and $X_2$ will increase $\Gamma_{1,2}$,  making it harder to tell whether a high value of $\hat{\Gamma}_{1,2}$ indicates that $\Gamma_{1,2}=1$ or that $\Gamma_{1,2}\lesssim 1$, as we shall see in Section~\ref{s:simul}.  

The discussion above suggests that conditioning on the values of known confounders might be valuable. In the presence of a vector $\mathbf{H}$ of potential confounders we therefore define 
\begin{equation}\label{e:popGammaCond}
\Gamma_{1,2\mid \mathbf{H}}:=\lim_{u\to 1^-}\mathbb{E}_{(X_1,X_2,\mathbf{H})}\left\{F_2(X_2\mid \mathbf{H})\mid F_1(X_1\mid \mathbf{H})>u\right\}.
\end{equation}
If there is no direct dependence of  $X_2$ on $X_1$, then $X_2$ is independent of $X_1$ conditional on $\mathbf{H}$, so $\Gamma_{1,2\mid \mathbf{H}}=1/2$, whereas $\Gamma_{1,2}$ lies in $ [1/2,1)$ but might be close to unity.  Thus $\Gamma_{1,2\mid \mathbf{H}}< \Gamma_{1,2}$ unless there are no confounders. If $X_1$ causes $X_2$, on the other hand, then $\Gamma_{1,2\mid \mathbf{H}}=\Gamma_{1,2}=1$. In the presence of potential confounders, therefore, \eqref{e:popGammaCond} seems preferable to $\Gamma_{1,2}$.  The difficulty is that the estimation of~\eqref{e:popGammaCond} requires the modelling of the dependence of both $X_1$ and $X_2$ on $\mathbf{H}$.  The first is more straightforward, because for large $u$ only the upper tail of $X_1$ need be considered, whereas the second ostensibly requires a model for the entire distribution of $X_2$, and this may be complex.  We compromise by fitting similar models to both variables, letting the upper tails alone vary with $\mathbf{H}$.  As we shall see below, this can greatly improve estimation of the causal dependence structure relative to the original approach.  Moreover fitting such a model should highlight simpler, potentially linear, structures in the tails, rather than more complex ones in the body of the data.  This leads us to propose a peaks-over-threshold approach to estimating the conditional dependence of $X_1$ and $X_2$ on $\mathbf{H}$ (Section~\ref{s:pctc}). Another useful tool, a reliable statistical test for direct causality, is discussed in Section~\ref{s:test}.

\section{Parametric Tail Causality and Confounder Dependence}\label{s:pctc}

\subsection{Generalized Pareto Causal Tail Coefficient}\label{ss:gpdctc}

As mentioned above, we use the generalized Pareto distribution (GPD) to model the tails of our variables~\citep[Chapter~4]{IExtr}.  For $j=1,2$, and under mild conditions on $X_j$, for a large enough threshold $u_j$ large enough, we have
\begin{equation}\label{eq:gpd}
\Prob(X_j-u_j\leq x\mid X_j>u_j)\approx G(x;\sigma_j,\xi_j)=1-\left(1+\xi_j x/\sigma_j\right)^{-1/\xi_j}_+,
\quad x>0, 
\end{equation}
with a {scale} parameter $\sigma_j>0$ and a {shape} parameter $\xi_j\in\Reals$:
\begin{itemize}
\item $\xi_j =0$ corresponds to light-tailed distributions, and then $X_j$ lies in the maximum domain of attraction of the Gumbel distribution;
\item $\xi_j >0$ corresponds to heavy-tailed  distributions, and then $X_j$ lies in the maximum domain of attraction of the Fr\'echet distribution; and
\item $\xi_j <0$ corresponds to distributions with bounded upper tails, and then $X_j$ lies in the maximum domain of attraction of the (reverse) Weibull distribution.
\end{itemize}
\indent
Any random variable satisfying the assumptions of Theorem~\ref{thm:ctc1val} satisfies~\eqref{eq:gpd}, as a regularly varying random variable with index $\alpha >0$ lies in the Fr\'echet maximum domain of attraction. If the threshold $u_j$ is chosen to be the $q$ quantile  of $X_j$ for some $q\in (0,1)$, then we can write 
\begin{equation*}
\begin{split}
     \Prob(X_j\leq x)  &\approx\left\{G(x-u_j;\sigma_j,\xi_j)(1-q)+q\right\}\1(x>u_j) + \Prob(X_j\leq x)\1(x\leq u_j), 
\end{split}
\end{equation*}
and using the empirical distribution $\hat F(x)$ to estimate $\Prob(X_j\leq x)$ and maximum likelihood estimation using the excesses of $u_j$ to obtain $\hat{\sigma}_j$ and $\hat{\xi}_j$ yields a hybrid estimator of the distribution function $F_j(x)$ of $X_j$, i.e., 
\begin{equation*}
\hat{F}_j(x;\hat{\sigma}_j,\hat{\xi}_j)=\hat{F}(x)\1(x\leq u_j)+\left\{G(x-u_j;\hat{\sigma}_j,\hat{\xi}_j)(1-q)+q\right\}\1(x>u_j).
\end{equation*}
The choice of $q$ involves a bias--variance trade-off: $q$ should be chosen large enough for the tail to be well approximated by a GPD, thus reducing the bias, but small enough to have enough exceedances, thus reducing the variance of the estimator.
Using hybrid estimators for $F_1$ and $F_2$ for an integer $k\in\{1,\ldots, n-1\}$ yields the parametric \emph{GPD causal tail coefficient} estimator for $\Gamma_{1,2}$,
\begin{equation}\label{eq:gpdctc}
    \hat{\Gamma}_{1,2}^{\rm GPD}=\dfrac{1}{k_g}\sum_{i=1}^n \hat{F}_2(X_{i,2};\hat{\sigma}_2,\hat{\xi}_2)\1\left\{\hat{F}_1(X_{i,1};\hat{\sigma}_1,\hat{\xi}_1)>1-k/n\right\},
\end{equation}%
where $k_g:=\lvert\{i\in\{1,\ldots,n\} : \; \hat{F}_1(X_{i,1};\hat{\sigma}_1,\hat{\xi}_1)>1-k/n \}\rvert$. Unlike with the non-parametric estimator~\eqref{eq:npctc1}, the number of data pairs $k_g$ used in~\eqref{eq:gpdctc} may not equal $k$, as it depends on the fit of $\hat{F}_1(X_{i,1};\hat{\sigma}_1,\hat{\xi}_1)$. %

The GPD model can be extended to allow dependence on covariates of interest by expressing its parameters in the form $\theta (i)=h\{\boldsymbol{\gamma}^\top\mathbf{Z}(i)\}$, where $\theta$ denotes one or both of $\sigma$ and $\xi$, $h$ is an inverse link function, $\boldsymbol{\gamma}$ is a vector of parameters and $\mathbf{Z}(i)$ is the vector of explanatory variables on which the model might depend~\citep{davison1990}.

We wish to reparametrise the model to reduce or remove the effect on $\Gamma_{1,2}$ of a vector of potential confounders $\mathbf{H}$ of $X_1$ and $X_2$. If $\mathbf{H}$ is part of the LSCM then under the setup in Section~\ref{s:ctc} it is straightforward to show that $\mathbf{H}$ affects the scale parameters of the GPD model that applies to $X_1$ and $X_2$ above high thresholds, but not their shapes, so we write
\begin{equation}
    \sigma_j(i) := \sigma_j^0+\boldsymbol{\sigma}_j^{1\top} \mathbf{H}_i, \quad i=1,\ldots,n,\; j=1,2,
\label{sigmaModel}
\end{equation}
where $\mathbf{H}_i$ is the replicate of $\mathbf{H}$ corresponding to the observations $(X_{i,1},X_{i,2})$ of $(X_1,X_2)$.

This yields, for $k\in\{1,\ldots, n-1\}$, the parametric \emph{$\mathbf{H}$-conditional linear generalized Pareto distribution (LGPD) causal tail coefficient estimator}, %
\begin{equation}\label{eq:lgpdctc}
    \hat{\Gamma}_{1,2\mid \mathbf{H}}^{\rm GPD}=\dfrac{1}{k_l}\sum_{i=1}^n \hat{F}_2\{X_{i,2};\hat{\sigma}_2(i),\hat{\xi}_2\}\1\left[\hat{F}_1\{X_{i,1};\hat{\sigma}_1(i),\hat{\xi}_1\}>1-k/n\right].
\end{equation}%
where $k_l:=\lvert\{i\in\{1,\ldots,n\} : \; \hat{F}_1\{X_{i,1};\hat{\sigma}_1(i),\hat{\xi}_1\}>1-k/n \}\rvert$.
Estimation of $\sigma_j^0$, $\sigma_j^1$ and $\xi_j$ is performed by maximum likelihood.
In applications it is preferable to center and rescale each confounder in $\mathbf{H}$ componentwise to unit variance and zero mean, to avoid numerical issues.
Although the confounder is here assumed to be part of the LSCM, this does not seem to be necessary in practice, as non-linear effects can be approximated linearly, especially in the tail region.  We investigate the effect of varying the tail index in Section~\ref{ss:difftail}.

\subsection{The Positive Linear Scale Issue}\label{sss:posscaleissue}

Linear modelling of the GPD scale parameter may not yield positive scale estimates $\hat{\sigma}_j(i)>0$ for each $i=1,\ldots,n$ and $j=1,2$. The use of a nonlinear link function to ensure that the scale estimates were positive would not agree with the assumption of extremal linearity of the causal relationships, as the effect of $\mathbf{H}$ on the scale is also necessarily linear. We now describe two different solutions to this problem, which we compare by simulation in Section~\ref{s:simul}.

The first solution, \emph{post-fit correction},\  replaces $\hat{\sigma}_j(i)$ in~\eqref{eq:lgpdctc} by $\max\{\hat{\sigma}_j(i),\epsilon \}$ for some arbitrary but small positive $\epsilon$. The second solution, the \emph{constrained approach},\ applies the following linear constraints to the estimates when maximizing the likelihood
\begin{equation}\label{eq:linbounds}
    \sigma_j^0+\boldsymbol{\sigma}_j^{1\top} \min_{i=1,\ldots,n} \mathbf{H}_i>0, \quad \sigma_j^0+\boldsymbol{\sigma}_j^{1\top} \max_{i=1,\ldots,n} \mathbf{H}_i>0, \quad j=1,2,
\end{equation}
where $\min_{i=1,\ldots,n} \mathbf{H}_i$ and $\max_{i=1,\ldots,n} \mathbf{H}_i$ represent the vectors of componentwise minima and maxima. When the data have a known distribution, box constraints can be used instead of~\eqref{eq:linbounds}.
For example, in the case of a single confounder $H$ and if $X_1$, $X_2$ and $X_h=H$  have $t_\nu$ distributions, then $\sigma_j^0=u_j/\nu$ and $\sigma_j^1=-\beta_{h\rightarrow i}/\nu$. Thus, if $ \sigma_j(i)=\sigma_j^0+\sigma_j^1 H_i>0$ $(j=1,2;i=1,\ldots,n)$, then 
\begin{equation}
-\dfrac{u_j}{\nu\max_{i=1,\ldots,n}H_i}<\sigma_j^1<-\dfrac{u_j}{\nu\min_{i=1,\ldots,n}H_i}, 
    \label{eq:studentbounds}
\end{equation}
where the lower and upper bounds are needed for positive and negative $H_i$, respectively.

\section{Simulation Study}\label{s:simul}%

Here we perform a simulation study using the Student $t$, Pareto and log-normal noise distributions. The first two lie in the Fr\'echet maximum domain of attraction and are regularly varying with index $\alpha=1/\xi>0$. We write ${\rm Pareto}(a,\alpha)$ for the Pareto model with scale parameter $a$ and tail index $\alpha$; recal that lower values of $\alpha$ indicate heavier tails. This distribution satisfies Definition~\ref{df:rvx} exactly, so one might expect Pareto data to show better behaviour than Student data. The log-normal distribution, ${\rm LogN}(\mu,\sigma^2)$  lies in the maximum domain of attraction of the Gumbel distribution and is not regularly varying, but finite samples from it can appear to be heavy-tailed.

\begin{figure}[t]
\centering
\includegraphics[width=0.8\textwidth]{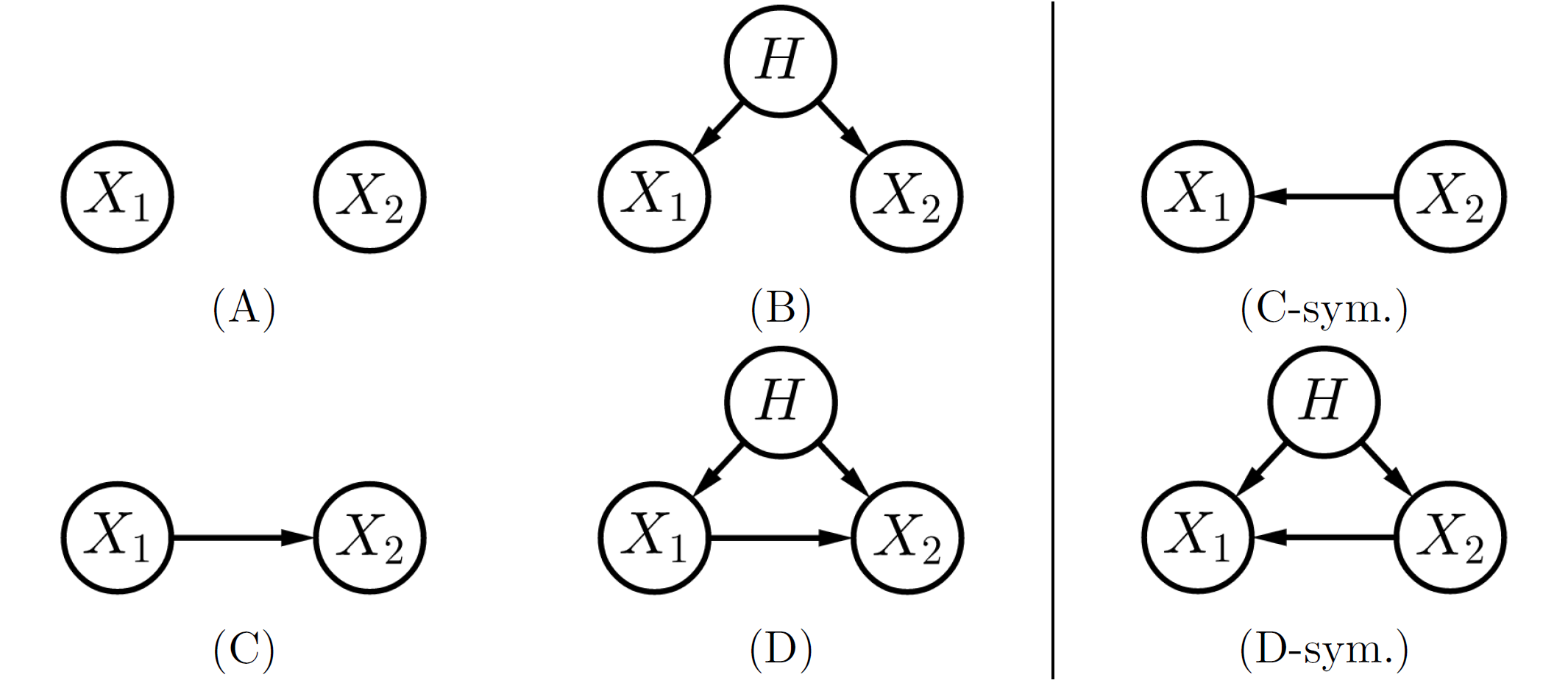}
\caption{The six possible causal configurations between $X_1$ and $X_2$ with a possible confounder $H$, separated into the four cases studied in the simulations, and the two omitted by symmetry.}
\label{f:causalstructs}
\end{figure}

We focus on the behaviour of the causal tail coefficient estimators~\eqref{eq:npctc1} and~\eqref{eq:lgpdctc} between two variables $X_1$ and $X_2$ in their causal configurations, as shown in Figure~\ref{f:causalstructs}. As we study the estimators of causal effects of both $X_1$ on $X_2$ and of $X_2$ on $X_1$, we generated simulations only for the four causal cases, A, B, C and D.
The LSCM causal weights $\beta_{2,1}$, $\beta_{1h}$ and $\beta_{2h}$ were chosen to equal $1.0$, by default, for each existing edge in all four cases. Hence, in D, $X_2$ is caused by $X_1$ and the single confounder $H$ with equal strength, even though $H$ has another effect on $X_2$ through $X_1$. 

Unless stated otherwise, each estimate is based on a random sample of $n=10^6$ triples $(X_1,X_2,H)$, of which $k=2\lfloor n^{0.4}\rfloor=502$ were chosen --- \citet{Gnecco2019} found that the optimal fractional exponent of $n$ for choosing $k$ seems to lie between $0.3$ and $0.4$. The factor $2$ doubles the number of data pairs used in the estimator, thus decreasing its variability, but does not introduce much bias for such large values of $n$. The GPD-based estimators are based on the top $(1-q)n$ observations, where we take $q=0.9$,  though only around $k$ of the largest observations are used to estimate the coefficients $\Gamma_{ij}$. Setting $q=0.95$ yields similar results. One thousand independent replicates were generated for each of the four causal configurations and three distributions. 

We present only the highlights of the study; the code and all the results are available from \href{https://github.com/opasche/ExtremalCausalModelling}{github.com/opasche/ExtremalCausalModelling}.

\subsection{Variables with Comparable Tails}\label{ss:const}

Detailed results for variables with comparable tails may be found in Section~\ref{sm:comparableTails} of the Supplementary Material. In this case it is essentially always possible to infer the existence and direction of any causality between $X_1$ and $X_2$, based on the non-parametric or $\mathbf{H}$-conditional LGPD estimators, \eqref{eq:npctc1} or~\eqref{eq:lgpdctc}, of $\Gamma_{1,2}$ and $\Gamma_{2,1}$ alone. When the causal effects of $H$ on $X_1$ and $X_2$, i.e., $\beta_{1h}$ and $\beta_{2h}$, are increased relative to the noise variance and any causal effect $\beta_{2,1}$ of $X_1$ on $X_2$, both $\Gamma_{1,2}$ and $\Gamma_{2,1}$ increase in configuration B, and $\Gamma_{2,1}$ increases in configurations C and D. This increase is larger with the non-parametric estimators of $\Gamma_{1,2}$ and $\Gamma_{2,1}$, which are biased upwards in these configurations. When the confounder has a high causal impact, inference based on the non-parametric estimator~\eqref{eq:npctc1} for direct causal link between $X_1$ and $X_2$ can fail, as $\hat{\Gamma}_{1,2},\hat{\Gamma}_{2,1}\approx 1$ and hence $\vert\hat\Gamma_{1,2}-\hat\Gamma_{2,1}\vert\approx 0$ in configurations B and D.

Use of the $\mathbf{H}$-conditional LGPD estimator~\eqref{eq:lgpdctc} greatly reduces the effect of $H$ on the coefficient estimates in configurations B and D. For Pareto and log-normal data, the results are indistinguishable from those without the confounder, both in terms of location and variability, as if the effect of $H$ had been entirely removed. The estimates based on Student data are also shifted to around the same values as in the corresponding confounder-free configurations, though their upper tails are marginally heavier. These few greater values remain appreciably lower than without $H$ as a covariate. For configurations A and C, unlike for B and D, the estimator is almost unaffected by the addition of $H$ as a covariate when it is not a confounder. This is also a useful property, as it could allow tests of whether a specific covariate is a confounder of two variables, based on changes to the estimated coefficients.

\subsection{Confounder with a Different Tail}\label{ss:difftail}
\begin{figure}[!t]
\centering
\includegraphics[width=0.93\textwidth]{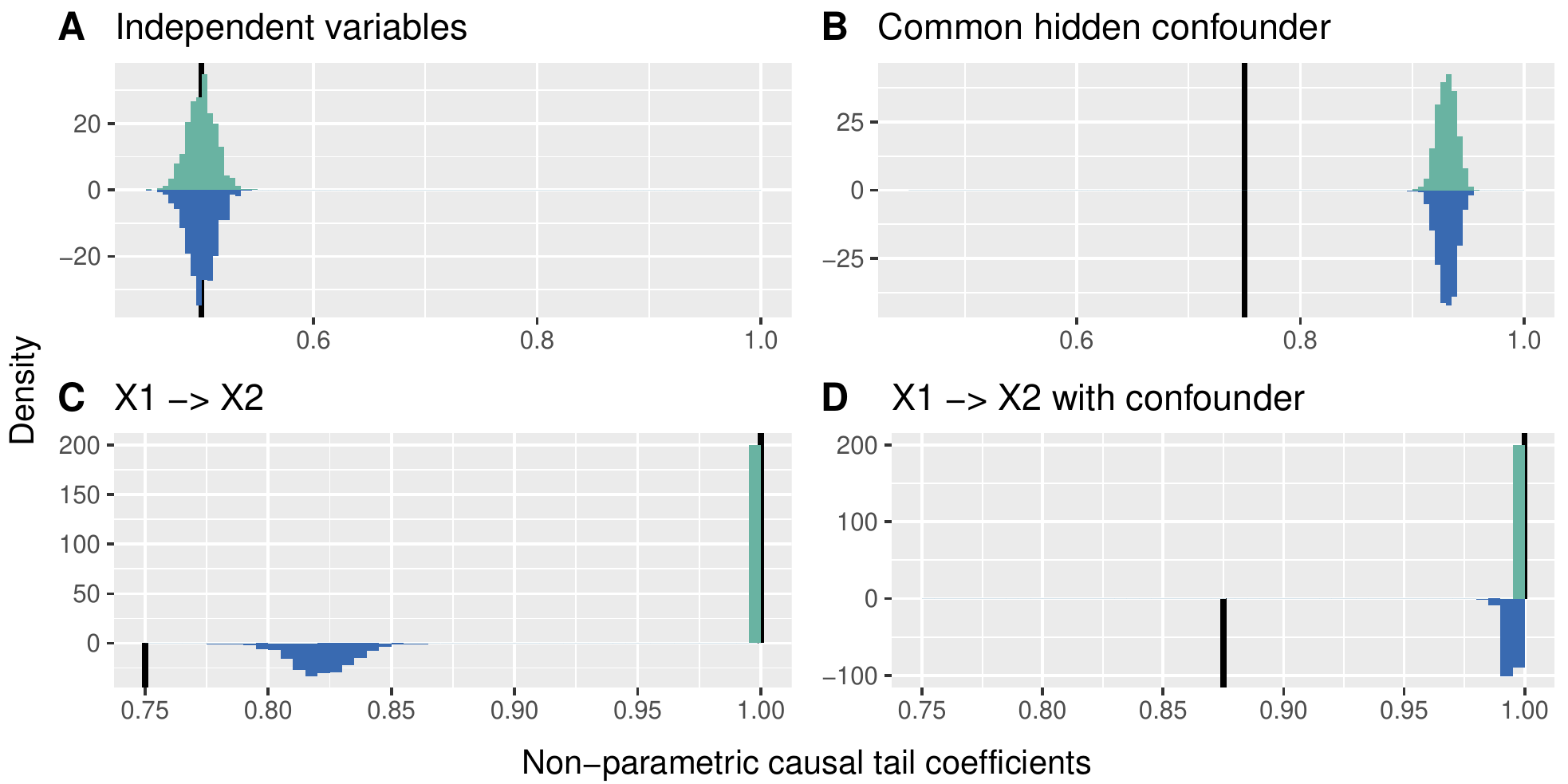}\\
\includegraphics[width=0.93\textwidth]{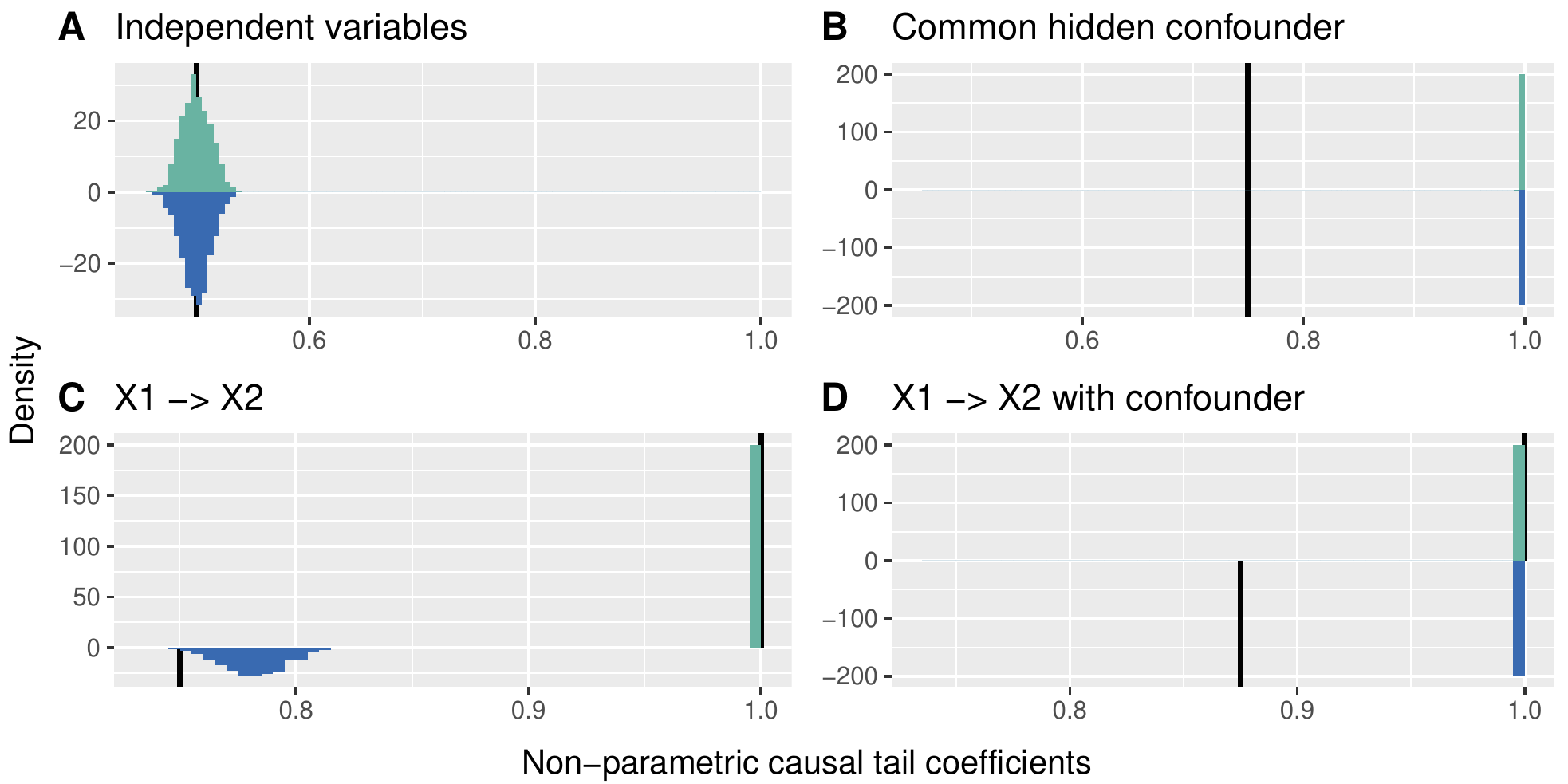}
\caption{Histograms of $\hat{\Gamma}_{1,2}$ (turquoise) and $\hat{\Gamma}_{2,1}$ (blue) for  $t_4$-distributed $\varepsilon_1$ and $\varepsilon_2$, and $t_3$-distributed $H$ (top four panels) and for ${\rm LogN}(0,1)$-distributed $\varepsilon_1$ and $\varepsilon_2$, and ${\rm LogN}(0,1.5)$-distributed $H$ (bottom four panels). Half-lines (black) indicate $\Gamma_{1,2}$ and $\Gamma_{2,1}$ for comparable tails.  
The panels for ${\rm Pareto}(1,3)$ distributed $\varepsilon_1$ and $\varepsilon_2$, and ${\rm Pareto}(1,1.5)$ distributed $H$ are very similar to the lower four panels.}
\label{fs:lnh1p5npctc}
\end{figure}

One generalisation allows the tail of the distribution of $H$ to be heavier or lighter than those of $X_1$ and $X_2$. A lighter tail does not negatively affect whether the non-parametric and $\mathbf{H}$-conditional LGPD estimators can infer a direct causal relationship between $X_1$ and $X_2$, as the tails of $X_1$ and $X_2$ then dominate.  
Figure~\ref{fs:lnh1p5npctc} shows the sampling distributions of $\hat{\Gamma}_{1,2}$ and $\hat{\Gamma}_{2,1}$ for all four causal structures when the tail of $H$ is heavier than those of $X_1$ and $X_2$. The true coefficient values are unknown, as assumption (b) of Theorem~\ref{thm:ctc1val} is not satisfied, though the coefficient for comparable tails,~\eqref{eq:lemctc1}, is shown for comparison.

When $H$ has a heavier tail than $X_1$ and $X_2$, the non-parametric estimators $\hat{\Gamma}_{1,2}$ and $\hat{\Gamma}_{2,1}$ in configuration B and $\hat{\Gamma}_{2,1}$ in configuration D are shifted well towards unity. With an even heavier-tailed, Student $t_2$,  distribution for $H$ (not shown here), the Student results resemble those for the Pareto and log-normal distributions. In all these cases it becomes impossible to infer a direct causal relationship between $X_1$ and $X_2$, owing to the effect of the heavier confounder tail on the non-parametric estimators.

\begin{figure}[tp]
\centering
\includegraphics[width=0.91\textwidth]{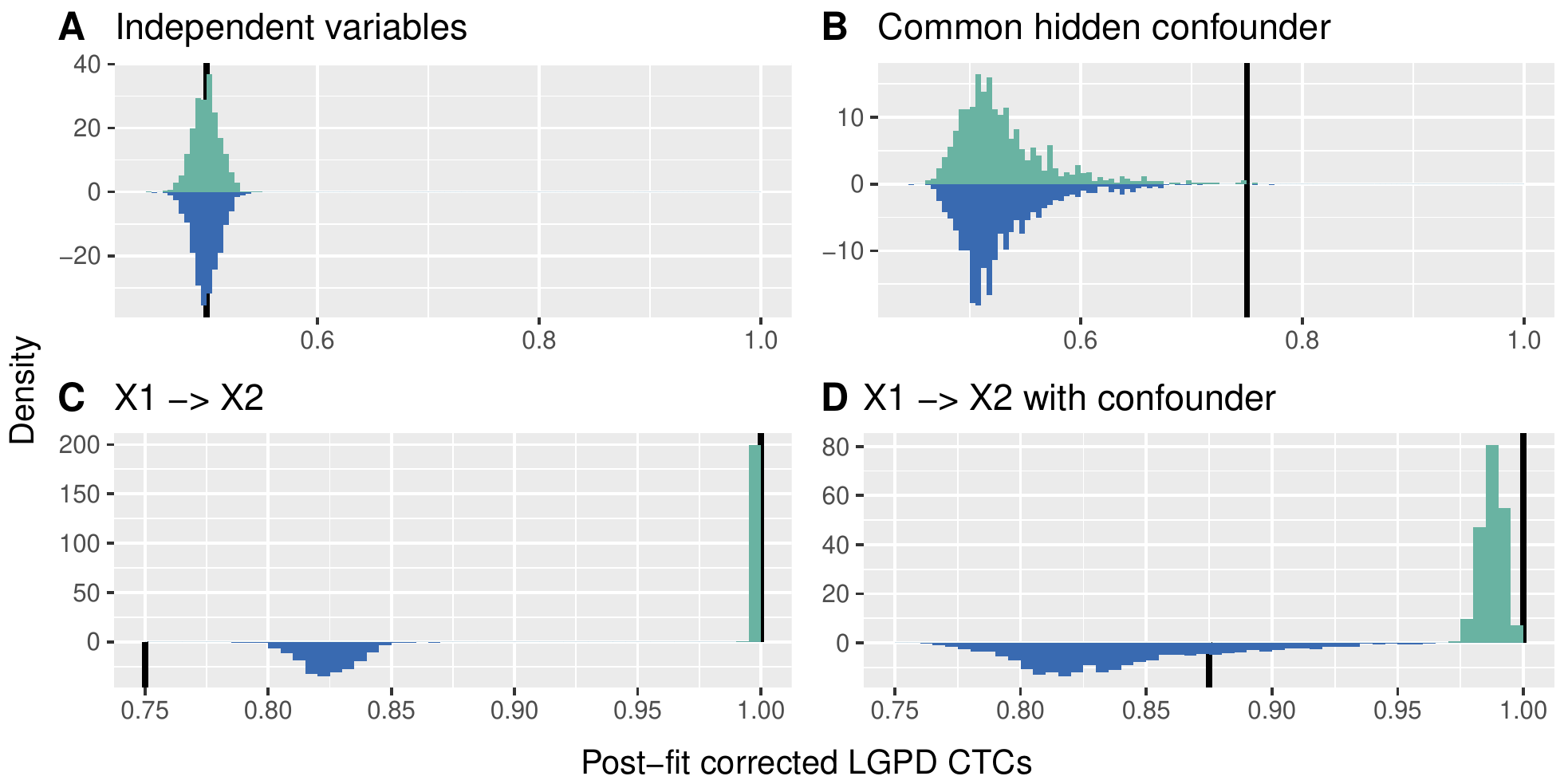}\\
\includegraphics[width=0.91\textwidth]{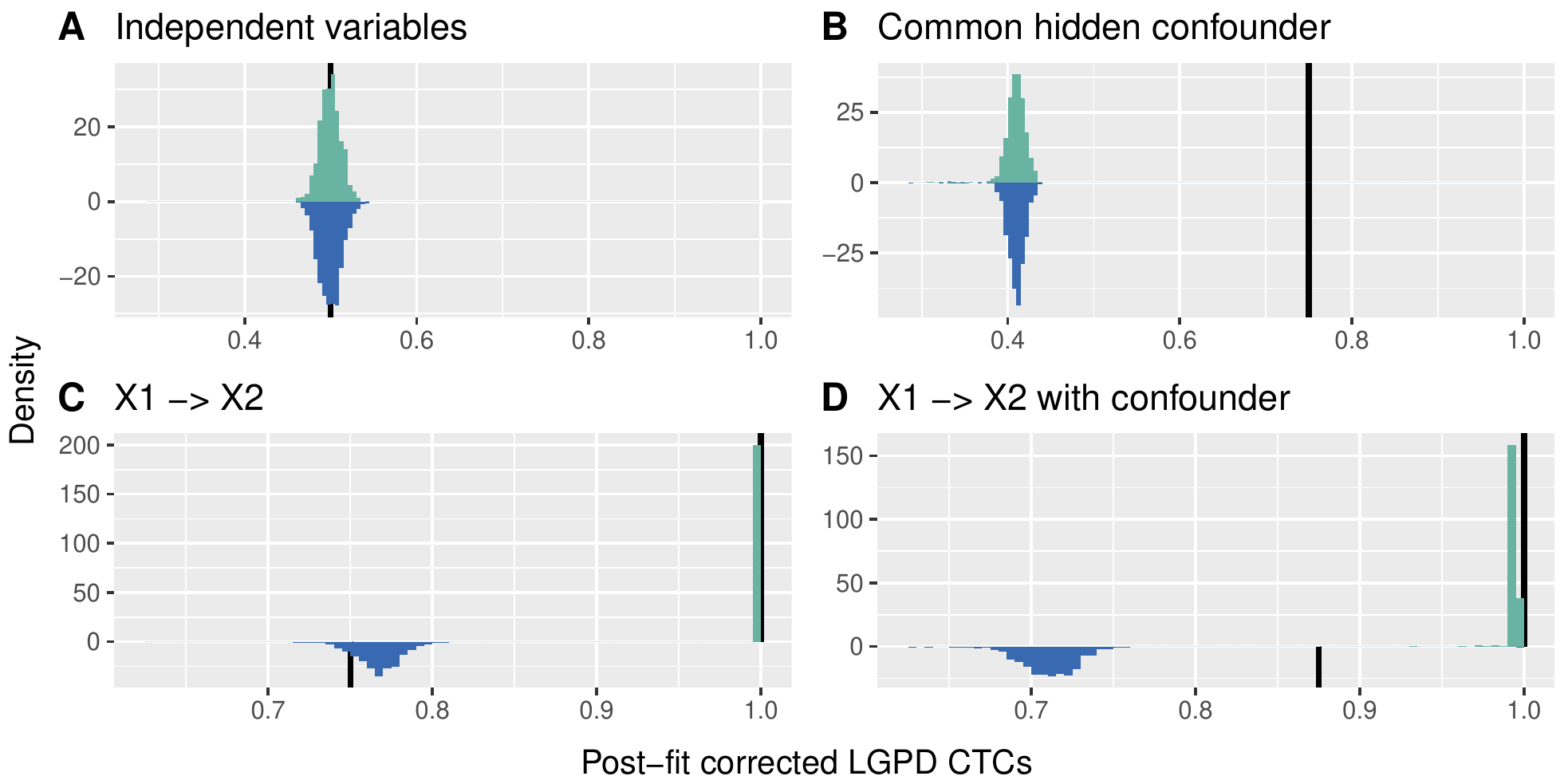}\\
\includegraphics[width=0.91\textwidth]{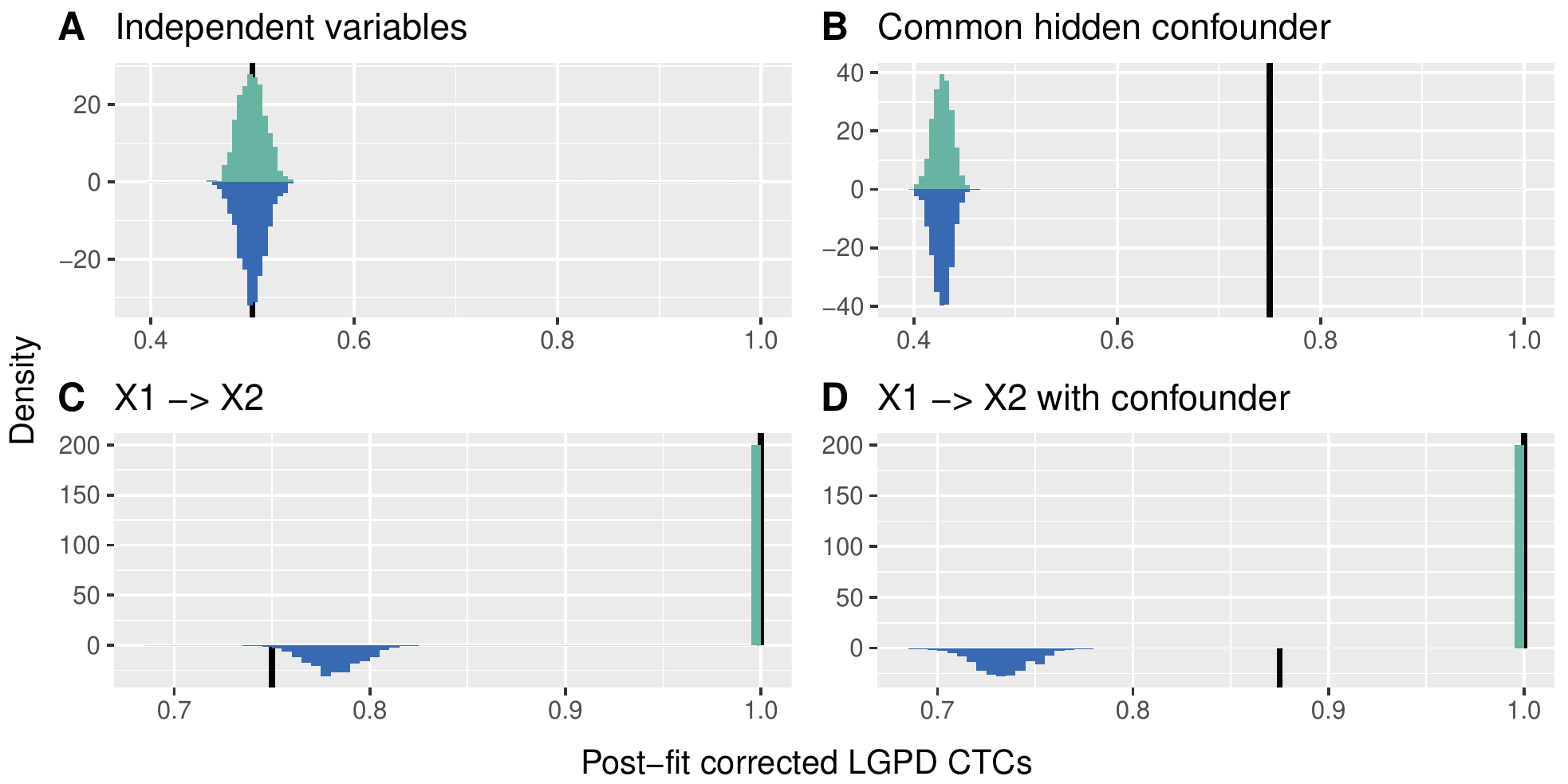}%
\caption{Histograms of $\hat{\Gamma}_{1,2\mid H}^{\rm GPD}$ (turquoise) and $\hat{\Gamma}_{2,1\mid H}^{\rm GPD}$ (blue) with post-fit correction for $t_4$ distributed $\varepsilon_1$ and $\varepsilon_2$, and $t_3$ distributed $H$ (top four panels),  for ${\rm Pareto}(1,3)$ distributed $\varepsilon_1$ and $\varepsilon_2$, and ${\rm Pareto}(1,1.5)$ distributed $H$ (middle four panels), and ${\rm LogN}(0,1)$ distributed $\varepsilon_1$ and $\varepsilon_2$, and ${\rm LogN}(0,1.5)$ distributed $H$ (lower four panels). Half-lines (black) indicate $\Gamma_{1,2}$ and $\Gamma_{2,1}$ for comparable tails.}
\label{fs:lnh1p5pflgpdctc}
\end{figure}

Figure~\ref{fs:lnh1p5npctc} shows that in configurations B and D the non-parametric estimator is badly affected by the heavier tail of $H$. %
Figure~\ref{fs:lnh1p5pflgpdctc}, which displays the sample distributions of $\hat{\Gamma}_{1,2\mid H}^{\rm GPD}$ and $\hat{\Gamma}_{2,1\mid H}^{\rm GPD}$ with post-fit correction when the tail of $H$ is heavier than those of $X_1$ and $X_2$,  shows that the use of $H$ as a covariate solves this problem: the estimates shift towards the coefficient values in the corresponding confounder-free cases, and consistently yield positive values of the difference  of estimates $\hat{\Gamma}_{1,2\mid H}^{\rm GPD} - \hat{\Gamma}_{2,1\mid H}^{\rm GPD}$ for configuration D and differences centred at zero for configuration B; see also Section~\ref{sm:comparableTails} of the Supplementary Material. The estimates in configurations A and C, without the confounder causal effect, are barely changed by using $H$ as a covariate.

\begin{figure}[!t]
\centering
\includegraphics[width=0.93\textwidth]{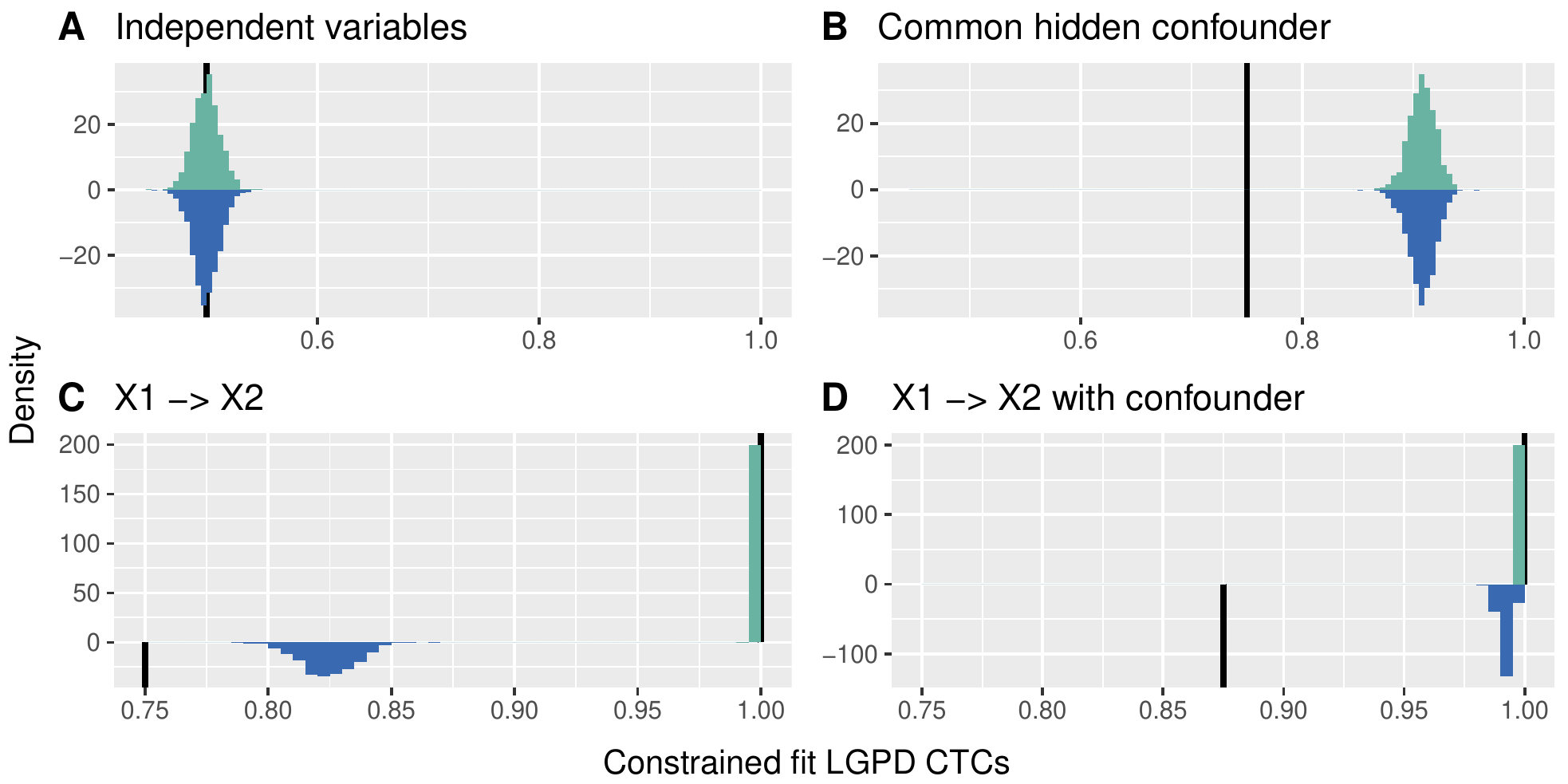}
\caption{Histograms of $\hat{\Gamma}_{1,2\mid H}^{\rm GPD}$ (turquoise) and $\hat{\Gamma}_{2,1\mid H}^{\rm GPD}$ (blue) with constrained fit for $t_4$ distributed $\varepsilon_1$ and $\varepsilon_2$, and $t_3$ distributed $H$. Half-lines (black) indicate $\Gamma_{1,2}$ and $\Gamma_{2,1}$ for comparable tails.}
\label{fs:t4h3cstrlgpdctc}
\end{figure}

Simulation results for $\hat{\Gamma}_{1,2\mid H}^{\rm GPD}$ and $\hat{\Gamma}_{2,1\mid H}^{\rm GPD}$ with the constrained fit are very similar to those for post-fit correction for the Pareto and log-normal distributions, but not for the Student distribution. Figure~\ref{fs:t4h3cstrlgpdctc} shows the sample distribution of $\hat{\Gamma}_{1,2\mid H}^{\rm GPD}$ and $\hat{\Gamma}_{2,1\mid H}^{\rm GPD}$ with the constrained fit, for a heavier confounder tail.
For the Student distribution, the confounder affects the estimator appreciably more for the constrained fit than for post-fit correction, compared to the non-parametric results. As the Student distribution is heavy in both tails, the lower constraint in~\eqref{eq:linbounds} forces $\hat{\sigma}_j(i)$ ($j=1,2$) to have an appreciably smaller slope,  explaining this reduced effect. In configurations with a confounder, the absolute values of the constrained $\hat{\sigma}_j^1$ may be up ten times smaller than those for post-fit correction. With both approaches $\hat{\sigma}_j^1$ rarely differs greatly from zero for configurations without a confounder.

Both types of constraint yield very similar estimates for the Student distribution; see \href{https://github.com/opasche/ExtremalCausalModelling}{github.com/opasche/ExtremalCausalModelling}.

To summarize, the simulations show that both the non-parametric estimator~\eqref{eq:npctc1} and the $\mathbf{H}$-conditional LGPD estimator~\eqref{eq:lgpdctc} perform well when the theoretical assumptions are met and the influence of a hidden confounder is limited.  When this influence grows, it becomes increasingly difficult to confidently infer the causal relationship between the variables using the non-parametric estimator, but the $\mathbf{H}$-conditional LGPD estimator allows us to detect this relationship by reducing the effect of the confounding. 

\vspace{2mm}

\section{Testing for Direct Causality}\label{s:test}

\subsection{Permutation Test}\label{ss:testexpl}

In situations such as the causal analysis presented in Section~\ref{s:rivers}, the distributions of the $\Gamma_{1,2}$ and $\Gamma_{2,1}$ estimators must be estimated to be used for inference. One way to obtain such distributions would be bootstrap resampling, but the extremal nature of the causal tail coefficient would require an unrealistically large sample size for its bootstrap  distributions to be trustworthy, as these distributions tend to be too discrete in the extremes.

We therefore propose a permutation test~\citep[Chapter~4]{DavisonBootstrap} for direct causality between two observed variables, measuring the asymmetry in their direct causal relationship. Suppose we have a sample $\left\{(X_{i,1},X_{i,2})\right\}_{i=1}^n$ from a LSCM and wish to test the null hypothesis of no direct causal relationship between $X_1$ and $X_2$,  $H_0: \beta_{2,1}=0$, versus the alternative that $X_1$ causes $X_2$, $H_A: \beta_{2,1}> 0$.  Our proposed procedure is as follows:
\begin{enumerate}
\item[1.] Rescale values $\tilde{X}_{i,j}=\tilde{F}_j(X_{i,j})$ ($i=1,\ldots,n$, $j=1,2$), where known confounders can be used in the distribution estimator $\tilde{F}_j$, as for $\hat{\Gamma}_{1,2\mid H}^{\rm GPD}$.
\item[2.] For $r=1,\ldots,R$, obtain $\tilde{X}_{i,1}^{(r)}$ and $\tilde{X}_{i,2}^{(r)}$ by randomly permuting the indices $j=1,2$ for each pair $(\tilde{X}_{i,1},\tilde{X}_{i,2})$ ($i=1,\ldots,n$).
\item[3.] Compute $\tilde{\Delta}_{1,2}=\tilde{\Gamma}_{1,2}-\tilde{\Gamma}_{2,1}$ on the transformed original data $\{(\tilde{X}_{i,1},\tilde{X}_{i,2})\}_{i=1}^n$ and $\tilde{\Delta}^{*r}_{1,2}=\tilde{\Gamma}_{1,2}^{*r}-\tilde{\Gamma}_{2,1}^{*r}$ on their bootstrapped values $\{(\tilde{X}_{i,1}^{(r)},\tilde{X}_{i,2}^{(r)})\}_{i=1}^n$ $(r=1,\ldots,R)$.
\item[4.] Obtain the Monte Carlo $p$-value, by comparing the value of the test statistic on the original rescaled data with the permutation distribution,
\begin{equation*}
p_{\rm mc}=\frac{1+\#_r\{\tilde{\Delta}^{*r}_{1,2}\geq\tilde{\Delta}_{1,2}\}}{R+1}.
\end{equation*}
\end{enumerate}

If there are no asymmetric confounding effects on the two variables, i.e.\ $\beta_{1h}=\beta_{2h}$ in the case of a single confounder, then $\Delta_{1,2}:=\Gamma_{1,2}-\Gamma_{2,1}=0$ under $H_0$, whereas $\Delta_{1,2}>0$ under $H_A$; see equation~\eqref{eq:lemctc1} and Theorem~\ref{thm:ctc1val}.
This does not hold generally with asymmetric confounding.
The direct causal relationship is symmetric under $H_0$, i.e., $X_2$ is as likely to take extreme values when $X_1$ is extreme as is $X_1$ when $X_2$ is extreme. If so, then permutations such as those performed in step 2.\ are equally likely, so $\tilde{\Delta}_{1,2}, \tilde{\Delta}^{*1}_{1,2},\ldots,\tilde{\Delta}^{*R}_{1,2}$ have a common distribution centered around zero, and $p_{\rm mc}$ will be uniformly distributed. Under the alternative, the direct causal relationship is ``asymmetric'', as $X_2$ is more likely to be extreme when $X_1$ is extreme than conversely; then $\tilde{\Delta}_{1,2}$ is more likely to lie in the upper tail of $\tilde{\Delta}^{*1}_{1,2},\ldots,\tilde{\Delta}^{*R}_{1,2}$. Thus the distribution of $p_{\rm mc}$ will become increasingly skewed towards zero as the causal strength of $X_1$ on $X_2$ increases.

If all asymmetric confounding effects are captured in $\tilde{F}_j$ by estimating the distribution conditionally, $X_1$ and $X_2$ have comparable tails and causal effects behave linearly in the extremes, then the proposed procedure should provide a reliable $p$-value for testing direct causality of $X_1$ on $X_2$.

\subsection{Simulations}\label{ss:testsims}

We used simulation from different data distributions and for different causal configurations involving  $X_1, X_2$ and a potential confounder $H$ to assess our proposed test. We used values of $0, 0.01, 0.05, 0.1, 0.2$ for the causal strength $\beta_{2,1}$ of $X_1$ on $X_2$, with confounding effects both present and absent. Symmetric ($\beta_{1H}=\beta_{2H}=1$) and asymmetric ($\beta_{1H}=0.8$ and $\beta_{2H}=1$, or $\beta_{1H}=1$ and $\beta_{2H}=0.8$) confounding effects were considered, and the noise variable were Pareto, Student $t$ and log-normal. We generated $m=10^3$ replicate samples of $n=10^4$ independent triples $(X_{i,1},X_{i,2},H_i)$ for each causal configuration and noise distribution. 
The sample size $n$ was chosen closer to practical orders of magnitude, compared to our large-sample study in Section~\ref{s:simul}.
Three versions of the permutation test were performed for each sample, corresponding to the causal tail coefficient estimators discussed in Sections~\ref{s:ctc} and~\ref{s:pctc}: the non-parametric~\eqref{eq:npctc1}, and $\mathbf{H}$-conditional LGPD~\eqref{eq:lgpdctc} with either post-fit correction or constrained fit. Each used $R=10^3$ permutations and the estimator hyper-parameters were set to $k=2\lfloor n^{0.4}\rfloor=78$ and $q=0.9$. 

\begin{figure}[!t]%
\centering
\includegraphics[width=\textwidth]{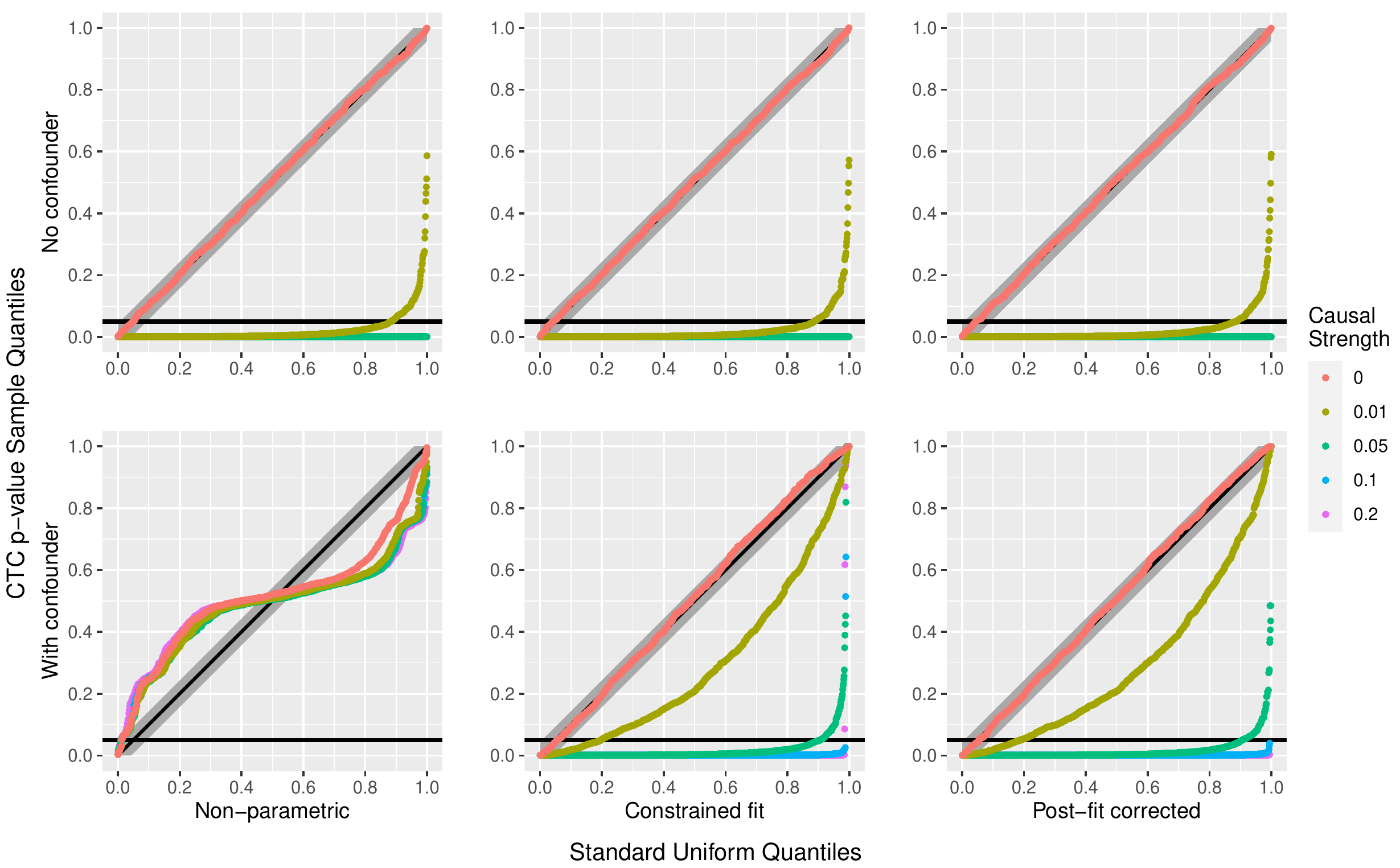}\\%
\includegraphics[width=\textwidth]{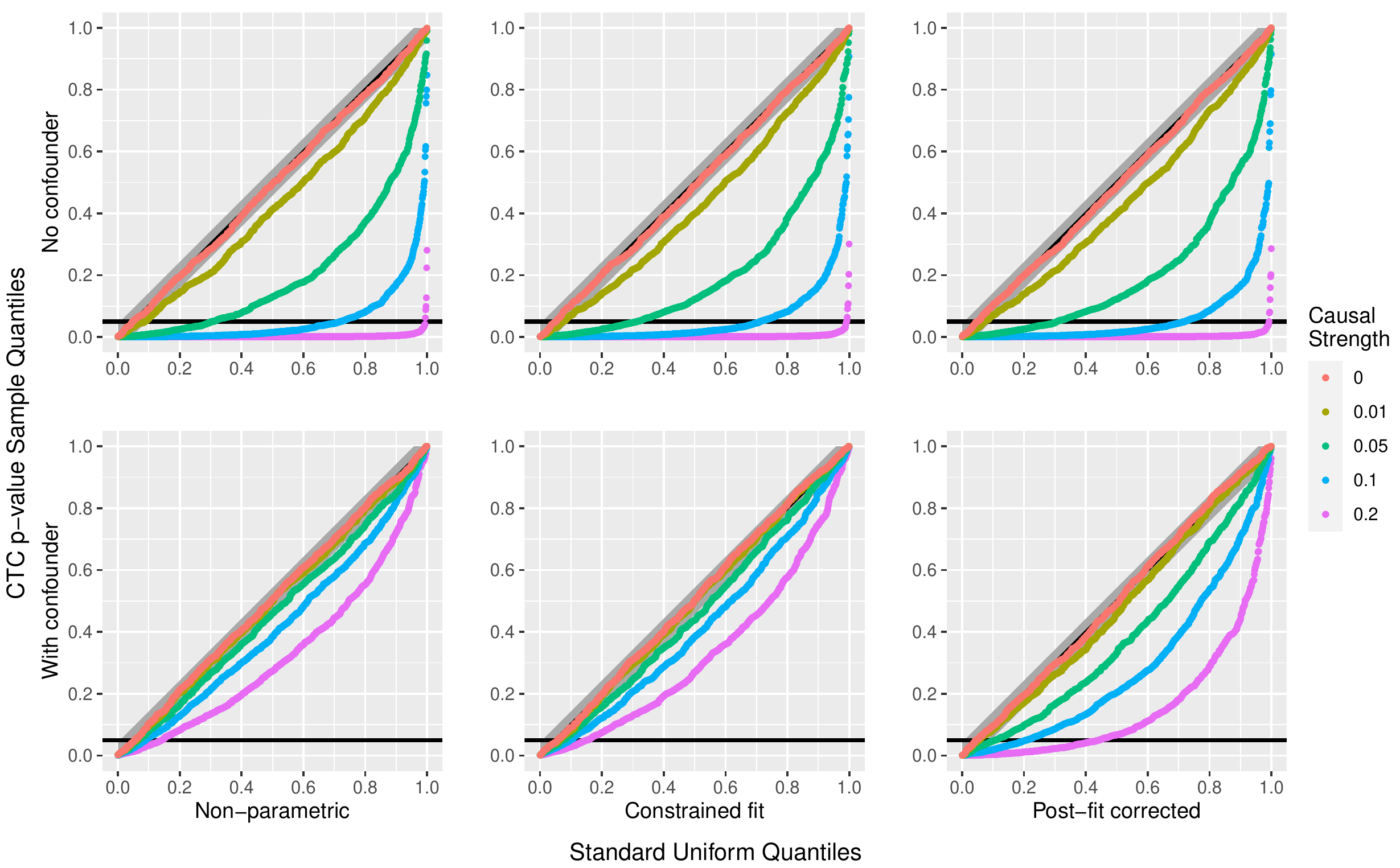}%
\caption{Uniform QQ-plots of Monte Carlo $p$-values $p_{\rm mc}$, with Kolmogorov--Smirnov confidence bands for different causal strengths $\beta_{2,1}$ (colors), the three estimators (columns) and optional symmetric confounding effects, $\beta_{1H}=\beta_{2H}=1$ (rows).  
Top six panels: ${\rm Pareto}(1,2)$ distributed $\varepsilon_1$ and $\varepsilon_2$, and ${\rm Pareto}(1,1)$ distributed $H$. 
Bottom six panels: $t_4$ distributed $\varepsilon_1$ and $\varepsilon_2$, and $t_3$ distributed $H$.}
\label{fs:QQtestPa2H1}
\end{figure}

Figure~\ref{fs:QQtestPa2H1} shows uniform QQ-plots of $p_{\rm mc}$ for the Pareto and Student distributions, in the case of heavier confounder tail, with symmetric effects. In the absence of confounding the test behaves as expected in both cases, and adding dependence on the independent $H$ variable in the modelling through the parametric estimators has no visible effect on the distribution of $p_{\rm mc}$ compared to the non-parametric approach. For the Pareto distribution, the test has a power of almost $0.9$ for a direct causal strength of $0.01$, and it behaves perfectly for higher causal strengths. For the Student distribution, the test reaches a power of $0.3$ for a direct causal strength of $0.05$, of $0.7$ for causal strength of $0.1$ and of near 1.0 for a causal strength of $0.2$.

When the confounding effects are added, the test based on the non-parametric estimator fails for the Pareto distribution, as most of the $p_{\rm mc}$ then lie outside  the $95\%$ confidence bands, indicating that the distribution of $p_{\rm mc}$ is highly non-uniform. This is corrected when the value of the confounder is taken into account using the parametric approaches, with power $0.9$ for a direct causal strength of only one twentieth of the confounder's marginal effects.  In the Student case, $p_{\rm mc}$ seems to be close to uniformity in the absence of direct causality (the difference in tail shape is much greater in the Pareto case), but  post-fit correction increases the power from below $0.2$ to above $0.4$ for a direct causal strength of one fifth of the confounder's marginal effects. Similar conclusions to those of Section~\ref{ss:difftail} about the constrained fit for distributions with both tails heavy apply, as the constrained fit estimator is not significantly better than the non-parametric estimator compared to post-fit correction.

\begin{figure}[t]
\centering
\includegraphics[width=\textwidth]{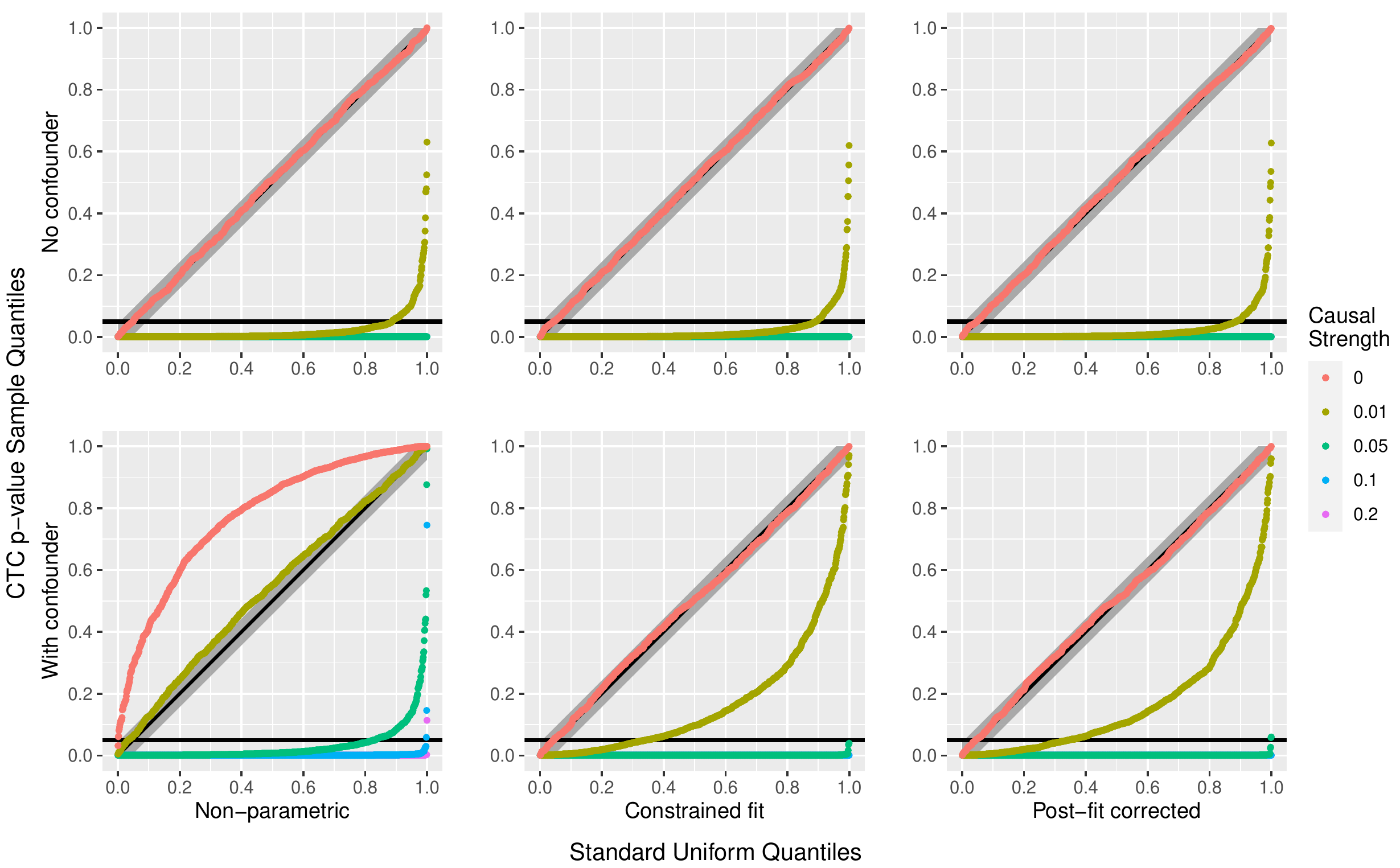}
\caption{QQ-plot of the $p_{\rm mc}$ estimates against the standard uniform distribution, with Kolmogorov--Smirnov confidence bands, for ${\rm Pareto}(1,2)$ distributed $\varepsilon_1$, $\varepsilon_2$ and $H$, for different causal strengths $\beta_{2,1}$ (colors), the three estimators (columns) and optional asymmetric confounding effects, $\beta_{1H}=0.8$, $\beta_{2H}=1$ (rows).}
\label{fs:QQtestPa2asym0p8_1}
\end{figure}

Figure~\ref{fs:QQtestPa2asym0p8_1} shows the uniform QQ-plots with asymmetric confounding effects for the Pareto distribution with comparable tails.
Unlike in the corresponding symmetric case, the test here fails when using the non-parametric estimator owing to the asymmetry induced by the confounder, but both parametric approaches remove this unwanted effect  by enough that $p_{\rm mc}$ nearly has a uniform distribution, with almost perfect power, for a causal strength of one sixteenth and one twentieth of the marginal confounding effects.

\section{Application to Swiss Rivers}\label{s:rivers}
We now illustrate how our method can discover direct causal relationships between the discharge extremes of pairs of river stations. This illustrates our method on a real example for which we know the `ground truth' of extremal causality, but unlike in the simulations of Section~\ref{s:simul}, we cannot control and do not know the true tail behaviour of the station discharges and their potential confounders. 

\subsection{Data Sources and Additional Collection}\label{sss:datasources}
We use the average daily discharges between January 1913 and December 2014 at the $68$ Swiss gauging stations shown in Figure~\ref{fd:chstations}, and add daily precipitation data from $105$ meteorological stations during the same period. Some additional information, such as the station elevation, catchment surface area and mean elevation, glaciation percent and coordinates, was collected from the Federal Office for the Environment's website. To reduce any seasonal effects due to unobserved confounders, we only consider data during June, July and August, as the more extreme observations happen during this period when mountain rivers are less likely to be frozen. Temporal clustering is likely to appear for average daily discharge data but can be captured by considering the average catchment precipitation as a covariate in the model for the GPD scale parameter \eqref{sigmaModel}.

\begin{figure}[!t]
\centering
\includegraphics[width=\textwidth]{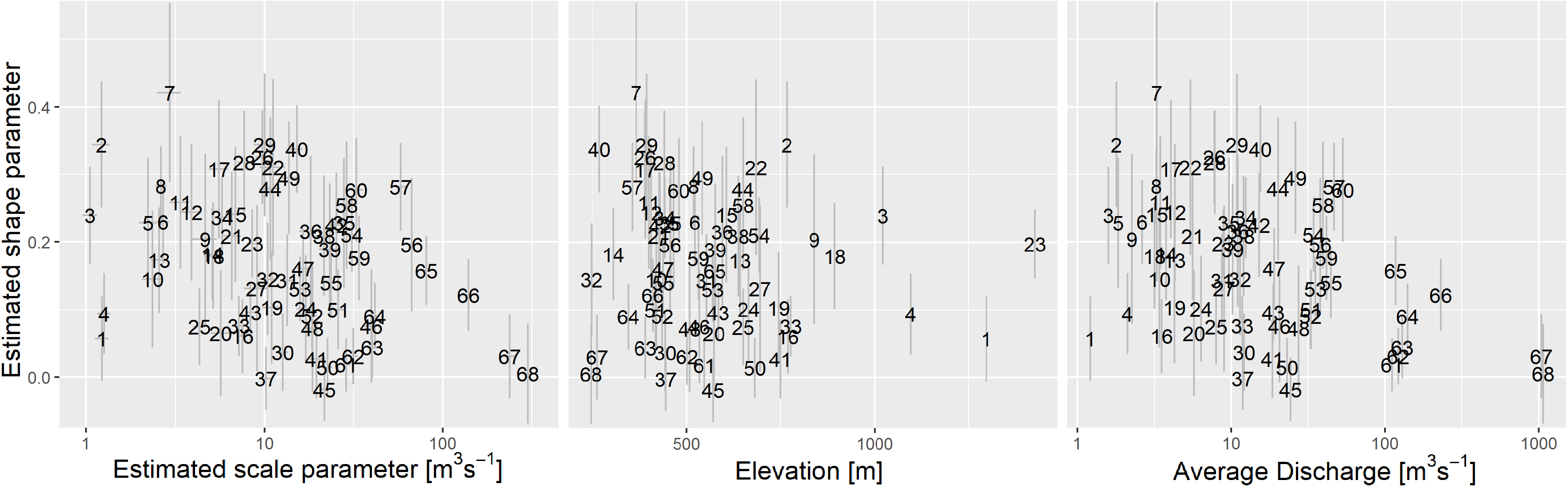}
\caption{Relation between shape parameter estimates, scale parameter estimates (log scale), station elevation and average discharge (log scale), with standard errors ($\pm {\rm SE}$) shown as error bars.}
\label{fd:shapech}
\end{figure}

Figure~\ref{fd:shapech} shows relationships between the estimates, station altitudes and average discharges. Altitude does not greatly affect the estimates, but the shape parameter estimates broadly decrease with increased average river discharge volume.

\subsection{Choice of Stations and Comonotonicity}\label{ss:choicestati}

For the causal analysis, we consider pairs of stations with known direct causal relationships, and pairs with no direct causal relationship. 
Causal pairs are ordered by the flow of water, with one downstream of the other. The river volumes for the pairs should be as similar as possible, as our exploratory analysis indicated different tail behaviours for rivers with very different average discharges. 
There should also be enough confluences between the two stations, otherwise one would observe \emph{comonotonicity}, i.e., almost perfect dependence,  between their discharges. If there is comonotonicity between $X_1$ and $X_2$, then $F_1(X_{i,1})\approx F_2(X_{i,2})$, for all $i=1,\ldots,n$, and it is impossible to know which variable causes which based on the data alone regardless of the approach, even if one is certain both of direct causality and of its direction. Confluences between the two stations reduce comonotonicity and make it possible to detect the direction of causality.

As we shall use precipitation as the confounding covariate, the stations must share likely meteorological effects and must lie in regions where precipitation data is available. Based on these criteria, we chose seven causal station pairs: $(43,62)$, $(42,63)$, $(36,63)$, $(24,61)$, $(44,61)$, $(22,38)$, $(22,35)$, where the first station of each pair lies upstream from the second.

The non-causal station pairs were selected to have similar average volume and similar shape parameter estimates. Pairs with stations separated by long distances and pairs relatively close to each other were both considered. The $13$  pairs selected are $(30,45)$, $(36,39)$, $(42,34)$, $(32,33)$, $(62,63)$, $(57,60)$, $(13,14)$, $(17,22)$, $(12,21)$, $(26,28)$, $(27,31)$, $(23,39)$, $(23,35)$.

The choice of covariate for the causal pairs was the mean daily precipitation among the meteorological stations in the area and the catchment of the two stations.
The choice of covariate was less meaningful for the non-causal pairs with large separating distances, which have different meteorological conditions, so the average daily precipitation over the whole country was used. For the pair $(42,34)$, which has the closest stations and local precipitation data available, the daily average in the local catchments was also considered. In the latter case, the pair will be highlighted with an asterisk to avoid confusion.

\subsection{Causal Analysis Results}\label{ss:statiresults}
For each station pair, the permutation test for direct causality was performed using the non-parametric~\eqref{eq:npctc1} and $\mathbf{H}$-conditional LGPD~\eqref{eq:lgpdctc} estimators with post-fit correction or constraints, with $R=10^4$ permutations and estimator hyper-parameters $k=1.5\lfloor n^{0.4}\rfloor$ and $q=0.9$. Table~\ref{t:swissresults} shows the values of $p_{\rm mc}$, the covariate shape estimate and its estimated extremal linear effects for the two stations, the latter estimated without constraints. The number of common observations for the pairs varies from $\numprint{2024}$ to $\numprint{8464}$, and $k$ lies between $31$ and $55$. With precipitation covariates added, the number of common observations ranges from $\numprint{1483}$ to $\numprint{7820}$, and $k$ lies between $27$ and $54$.

\begin{table}[t]
\centering
\caption{Permutation $p$-values $p_{\rm mc}$ for station pairs using the non-parametric approach (NP), the $\mathbf{H}$-conditional post-fit corrected (PFC) and constrained fit (CF) LGPD approaches, and an $\mathbf{H}$-conditional exponential inverse-link GPD approach (Exp). The shape estimate $\hat{\xi}_H$ for the precipitation covariate and the unconstrained scale slope estimates are also shown (with standard errors of at most $0.03$ for the former and in parentheses for the latter).\label{t:swissresults}}
\nprounddigits{2}
\begin{tabular}{ll|n{1}{2}|n{1}{2}|n{1}{2}|n{1}{2}|n{1}{2}|n{1}{2}|n{2}{2}|}
Stations & Pair type & \text{NP}  & \text{PFC} & \text{CF} & \text{Exp} & \text{$\hat{\xi}_H$} & \text{$\hat{\sigma}_{1}^{1}$} & \text{$\hat{\sigma}_{2}^{1}$} \\
\hline
43-62 & causal    & 0.0080991901 & 0.0079992001 & 0.0074992501 & 0.0080991901 & 0.0566669337 & 0.8842009468 (0.3) & 1.9066945453 (1.3) \\
42-63 & causal    & 0.0279972003 & 0.0211978802 & 0.0245975402 & 0.0351964804 & 0.0601108523 & 6.4878543848 (1.1) & 8.6026410091 (2.2) \\
36-63 & causal    & 0.0274972503 & 0.0226977302 & 0.0198980102 & 0.0295970403 & 0.0601108523 & 5.027641044  (1.1) & 7.2532593647 (2.8) \\
24-61 & causal    & 0.0555944406 & 0.0074992501 & 0.0053994601 & 0.0039996    & -0.014601569 & 3.4157640955 (1.2) & -2.3350882618(2.4) \\
44-61 & causal    & 0.0104989501 & 0.00449955   & 0.00459954   & 0.0065993401 & 0.0127824624 & 1.8939044269 (0.7) & -1.2100340565(2.0) \\
22-38 & causal    & 0.5802419758 & 0.402859714  & 0.397660234  & 0.3305669433 & 0.0659749896 & 3.425892068  (0.8) & 8.000950484  (2.0) \\
22-35 & causal    & 0.2177782222 & 0.1730826917 & 0.1745825417 & 0.097190281  & 0.0295179206 & 3.4292170544 (0.9) & 11.666130149 (3.0) \\\hline
30-45 & non-caus. & 0.5578442156 & 0.4658534147 & 0.4668533147 & 0.4561543846 & 0.0052839693 & 1.0132876404 (0.4) & 0.8868363734 (0.9) \\
36-39 & non-caus. & 0.796620338  & 0.699030097  & 0.698330167  & 0.6874312569 & 0.0052839693 & 4.6059105148 (1.1) & 4.168906454  (1.6) \\
42-34 & non-caus. & 0.2264773523 & 0.0422957704 & 0.0408959104 & 0.099690031  & 0.0052839693 & 5.967653909  (1.2) & 0.4274070406 (0.3) \\
42-34$^*$ & non-caus. & 0.2264773523 & 0.1303869613 & 0.1250874913 & 0.1125887411 & 0.0543840826 & 6.2882944671 (1.1) & 0.6556023806 (0.3) \\
32-33 & non-caus. & 0.0098990101 & 0.0096990301 & 0.0099990001 & 0.00159984   & 0.0052839693 & 0.6325946323 (0.4) & 1.0012907206 (0.3) \\
62-63 & non-caus. & 0.101389861  & 0.4889511049 & 0.4753524648 & 0.299970003  & 0.0052839693 & 1.08066753   (1.4) & 7.6742023274 (2.1) \\
57-60 & non-caus. & 0.9923007699 & 0.99930007   & 0.99910009   & 0.9990001    & 0.0052839693 & 6.3100949401 (3.7) & 5.2283442628 (1.8) \\
13-14 & non-caus. & 0.3215678432 & 0.5590440956 & 0.5639436056 & 0.5337466253 & 0.0052839693 & 0.5925547817 (0.2) & 1.1850691043 (0.3) \\
17-22 & non-caus. & 0.0050994901 & 0.0547945205 & 0.0565943406 & 0.0478952105 & 0.0052839693 & 0.776163342  (0.5) & 2.1809668187 (0.7) \\
12-21 & non-caus. & 0.5146485351 & 0.498550145  & 0.500349965  & 0.7209279072 & 0.0052839693 & 0.7126162033 (0.3) & 1.3336258551 (0.4) \\
26-28 & non-caus. & 0.6345365463 & 0.900109989  & 0.8927107289 & 0.9220077992 & 0.0052839693 & 1.9041701392 (0.5) & 1.6324917792 (0.4) \\
27-31 & non-caus. & 0.404459554  & 0.6259374063 & 0.6215378462 & 0.7521247875 & 0.0052839693 & 1.7129163243 (0.7) & 2.9055730916 (1.1) \\
23-39 & non-caus. & 0.802719728  & 0.9134086591 & 0.9184081592 & 0.9323067693 & 0.0052839693 & 2.4975701839 (0.6) & 4.2681161898 (1.5) \\
23-35 & non-caus. & 0.6484351565 & 0.8844115588 & 0.8877112289 & 0.8647135286 & 0.0052839693 & 2.4975701839 (0.6) & 6.6610856116 (1.7) \\
\hline
\end{tabular}
\npnoround
\end{table}

With the non-parametric approach for the causal stations, the absence of direct causality was rejected for four of the seven station pairs at significance level $5\%$, and for two of these four at level $2.5\%$. Adding  daily precipitation as a covariate by either parametric approach decreases the $p$-values but two pairs remain non-significant; both lie in the same region and contain station~$22$.

With the non-parametric approach, the absence of direct causality was not rejected for ten of the $13$ non-causal station pairs. Adding precipitation as a covariate with the two parametric approaches `corrected' the p-value for another station. For the pair $(42,34)$ using local instead of global precipitation as a covariate gave a higher p-value.

We also considered using an exponential rather than a linear inverse-link function, i.e., taking $\log \sigma_j(i) = \sigma_j^0+\sigma_j^1 H_i$ $(i=1,\ldots,n;j=1,2)$, to avoid any need for correction or constraints. The resulting $p_{\rm mc}$ values, also shown in Table~\ref{t:swissresults}, lead to the same conclusions as with the linear approaches.

Using the usual normal approximation, every $\hat{\sigma}_{1}^{1}$ is significantly positive for the causal pairs and $10$ of the $14$ estimates are positive for the non-causal pairs, with the highest confidence for the pair using local precipitation. Standard errors for $\hat{\sigma}_{2}^{1}$ are systematically larger than those for $\hat{\sigma}_{1}^{1}$ for the causal pairs, perhaps owing to the double causal effect of the covariate on the downstream station, both direct and indirect through the upstream station, as we do not observe this systematically for non-causal pairs. Consequently, the $\hat{\sigma}_{1}^{1}$ estimates are significantly positive for only four of the seven causal pairs, to be contrasted with $12$ of the $14$ estimates for the non-causal pairs. In particular, only the local precipitation effect is significant for the pair $(42,34)$.

We compare our results to two classical causal inference approaches appropriate to our problem. These are a non-Gaussian method for estimating causal linear structures based on results from independent component analysis, ICA-LiNGAM~\citep{Shimizu06}, and the PC algorithm, which retrieves the completed partially directed acyclic graph by performing conditional independence tests on the variables. For the latter, we consider both the classic PC algorithm \citep{spirtes00}, which uses Gaussian conditional independence tests, and the Rank PC algorithm~\citep{harris13}, which uses rank-based Spearman correlation to perform the independence tests and thus is more robust to non-normal variables. The results for the ICA-LiNGAM method are presented in Table~\ref{t:swisscompet} in the Supplementary Material, which shows the linear causal coefficients for the discharge station pairs estimated with the ICA-LiNGAM algorithm using either the station pair only (two variables) or the station pair and precipitation (three variables). Non-null values indicate significant causal effects. The upper-script arrows indicate the estimated direct causal direction between the station pair. Although in both cases of the two or three variables, ICA-LiNGAM retrieves all the correct causal pairs, with correct direction, all the non-causal pairs are indicated by non-null values as significantly causal. Both versions of the PC algorithm, once applied to our 21 pairs, provide existing direct causal links (without weights nor direction) between all the pairs of stations. Apparently both ICA-LiNGAM and PC methods are too eager to detect causality, unlike the tail coefficients. One explanation could be a set of unobserved confounders related to common global weather conditions triggering causal effects even between stations that are far apart. Extreme discharges depend more on local weather conditions, and particularly on heavy precipitation.
Another explanation could be that causal effects are only linear in the tails, perhaps due to ground saturation by precipitation.

\section{Discussion and Conclusion}
\label{s:conclu}
This paper addresses the reduction or removal of the unwanted effect of known confounders from the extremal causal analysis between two variables and the discovery of extremal causal relationships using a parametric estimator of the causal tail coefficient, based on generalized Pareto modelling, and a permutation test for direct causality.  Both allow the use of known confounders as covariates.

In our simulation study, the new estimator removed the confounder's unwanted effect almost entirely for variables with comparable tails, and reduced its effect enough to allow correct causal inference on the direct causal relationship in the case of a confounder with a heavier tail. The permutation test was shown to provide reliable $p$-values when all asymmetric confounding effects are captured in the model.

When applied to Swiss river discharge data, our methodology allowed correct inference on the direct causal relationships between discharges for the majority of the chosen station pairs, and the parametric approach captured the confounding effect of precipitation.

In many real-life situations, statistically significant covariates need not correspond to causal effects. \citet{CausalPred} have proposed a methodology for causal discovery, for when data  from different settings or regimes are observed.   Their method constructs invariant causal regression or classification models that should still make accurate predictions under interventions on the covariates or a change of environment. Adapting this approach to our setting would lead to a better understanding of causality of extremes.

\section*{Acknowledgements}
The work was supported by the Swiss National Science Foundation.

\bibliographystyle{abbrvnat-namefirst}%
\bibliography{OCP_bibliography}

\newpage

\begin{appendix}

\thispagestyle{empty}

\vbox{%
\hsize\textwidth
\linewidth\hsize
\vskip 0.1in
\centering
{\Large\bf SUPPLEMENTARY MATERIAL TO \\
\vspace{12pt}
``Causal Modelling of Heavy-Tailed Variables and Confounders with Application to River Flow''}\\

\textsc{}\\ %
\vskip 0.1in
\begin{tabular}[t]{c}\bf\rule{0pt}{24pt}%
Olivier~C.~Pasche\\
Research Center for Statistics, University of Geneva, Switzerland,\\
Institute of Mathematics, EPFL, 1015 Lausanne, Switzerland,\\
\href{mailto:olivier.pasche@unige.ch}{\texttt{olivier.pasche@unige.ch}}\\
\end{tabular}\hfil\linebreak[0]\hfil%
\begin{tabular}[t]{c}\bf\rule{0pt}{24pt}\ignorespaces%
Val\'erie~Chavez-Demoulin\\
Faculty of Business and Economics, University of Lausanne, Switzerland,\\
\href{mailto:valerie.chavez@unil.ch}{\texttt{valerie.chavez@unil.ch}}\\
\end{tabular}\hfil\linebreak[0]\hfil%
\begin{tabular}[t]{c}\bf\rule{0pt}{24pt}\ignorespaces%
Anthony~C.~Davison\\
Institute of Mathematics, EPFL, 1015 Lausanne, Switzerland,\\
\href{mailto:anthony.davison@epfl.ch}{\texttt{anthony.davison@epfl.ch}}\\
\end{tabular}%
\vskip 0.4in
\vskip 0.2in
}

\setcounter{section}{0}
\setcounter{subsection}{0}
\setcounter{equation}{0}
\setcounter{figure}{0}
\setcounter{table}{0}
\renewcommand{\thesection}{S.\arabic{section}}
\renewcommand{\thesubsection}{S.\arabic{section}.\arabic{subsection}}
\renewcommand{\theequation}{S.\arabic{equation}}

\renewcommand\theHsubsection{S.\thesubsection}
\renewcommand\theHequation{S.\theequation}
\renewcommand\thefigure{S.\arabic{figure}}
\renewcommand\thetable{S.\arabic{table}}

\section{Variables with Comparable Tails}\label{sm:comparableTails}
\subsection{Non-Parametric Causal Tail Coefficient Estimator}
\begin{figure}[p]
\centering
\includegraphics[width=0.93\textwidth]{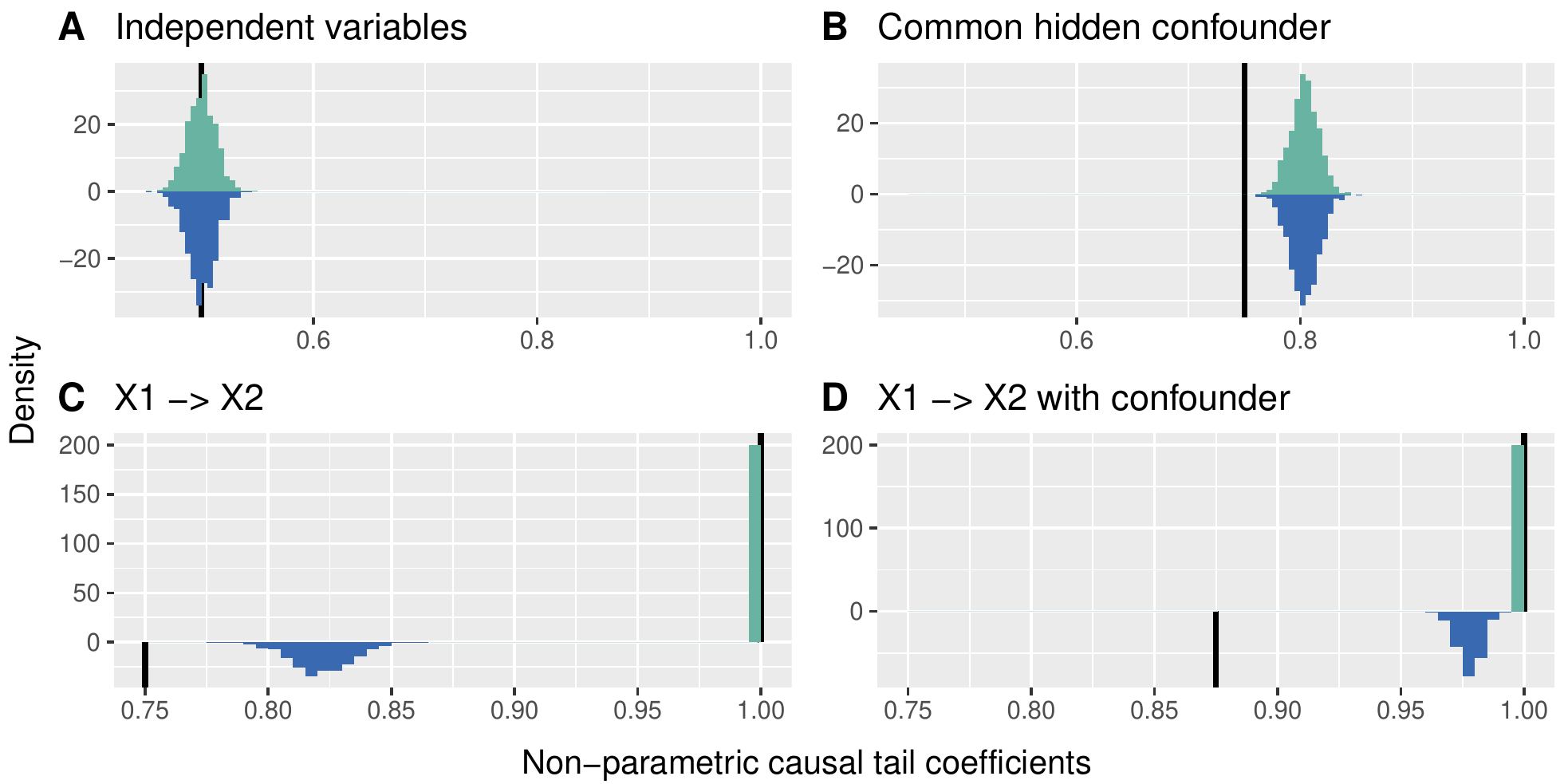}\\%
\includegraphics[width=0.93\textwidth]{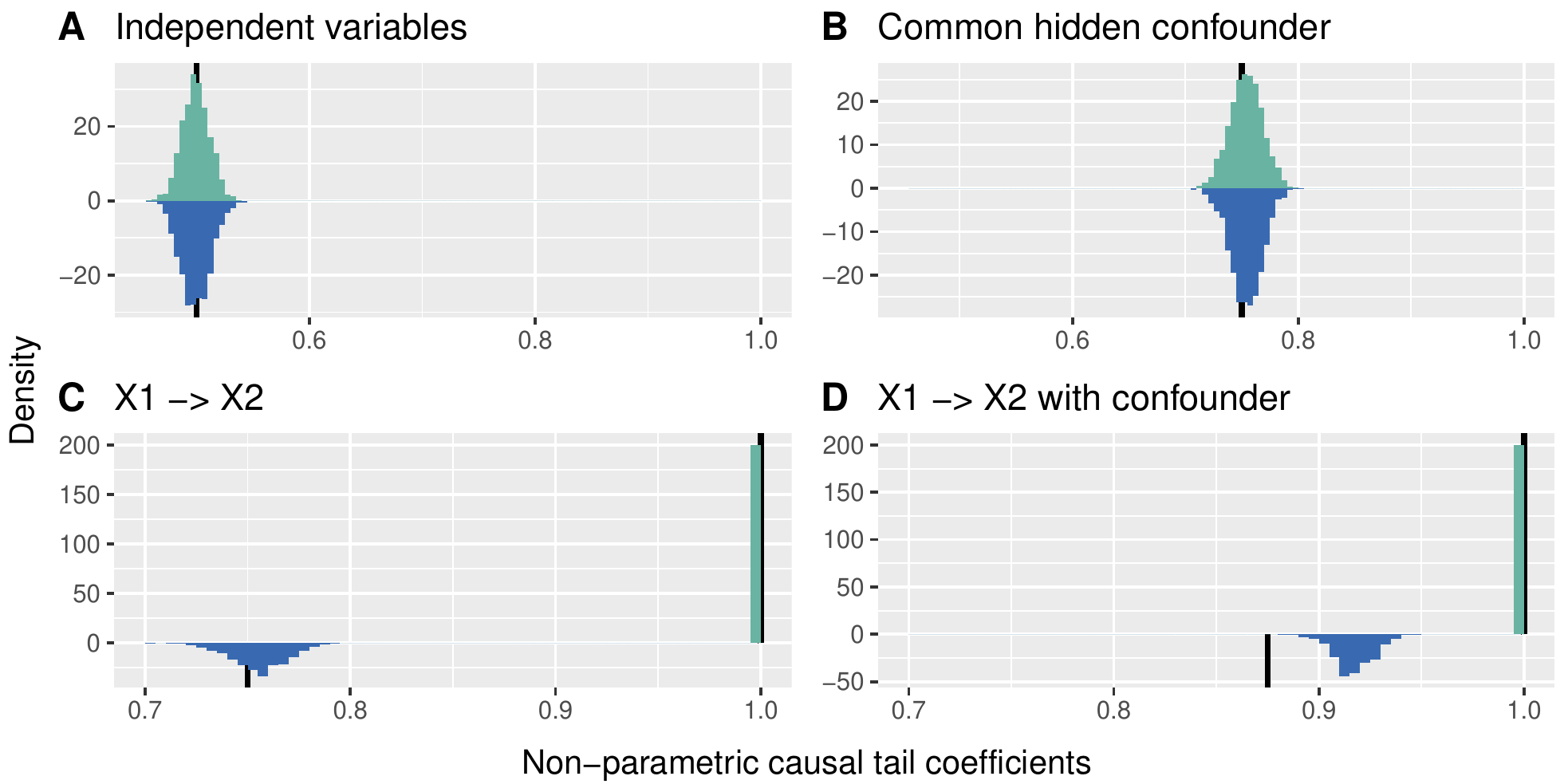}\\%
\includegraphics[width=0.93\textwidth]{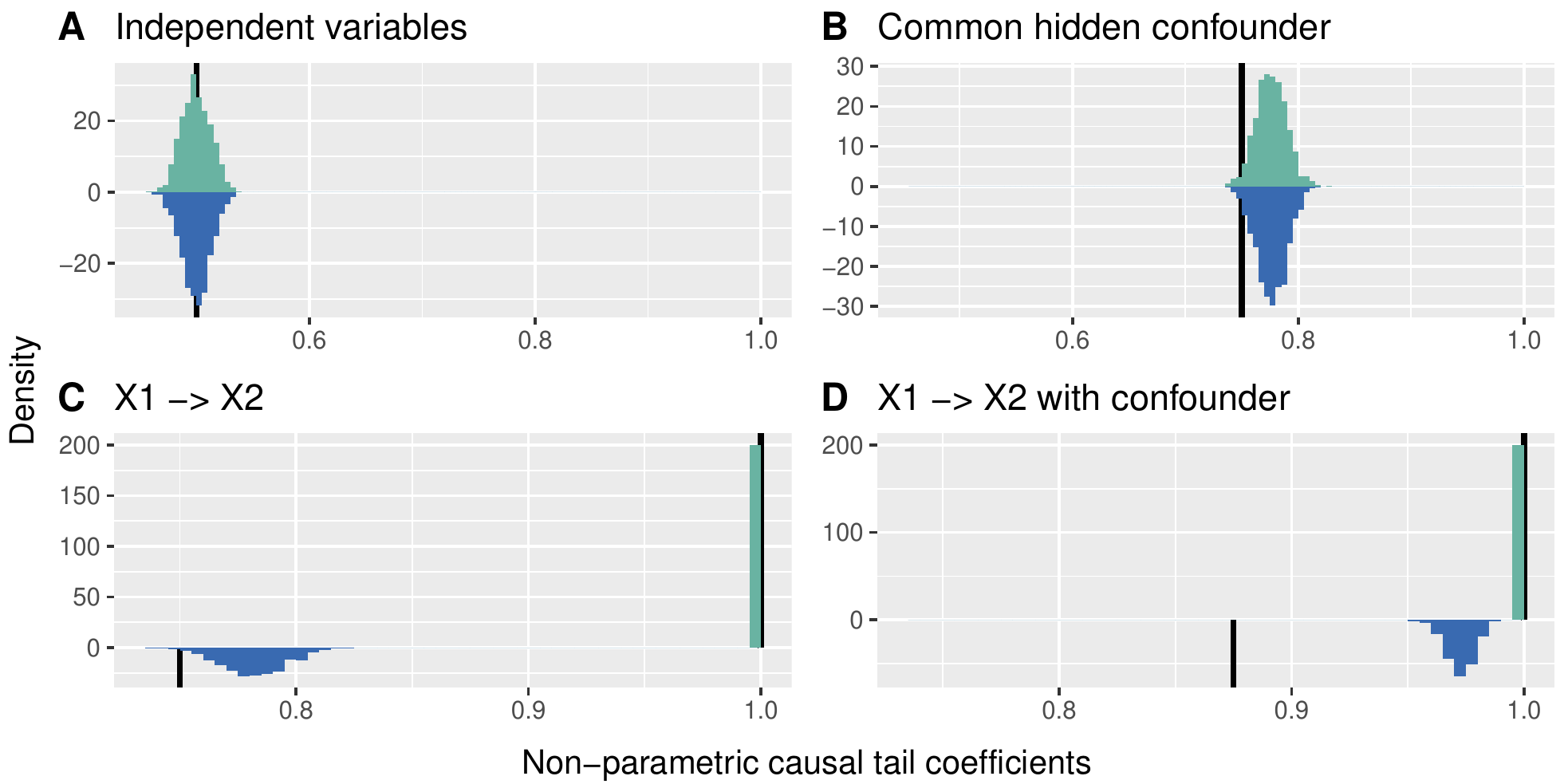}%
\caption{Histograms of $\hat{\Gamma}_{1,2}$ (turquoise) and $\hat{\Gamma}_{2,1}$ (blue) for $t_4$ (top four panels), ${\rm Pareto}(1,2)$ (middle four panels) and ${\rm LogN}(0,1)$ (bottom four panels) distributed noise variables, for the four causal configurations. Half-lines (black) indicate $\Gamma_{1,2}$ and $\Gamma_{2,1}$.}
\label{fs:npctc}
\end{figure}

\begin{figure}[p]
\centering
\includegraphics[width=0.93\textwidth]{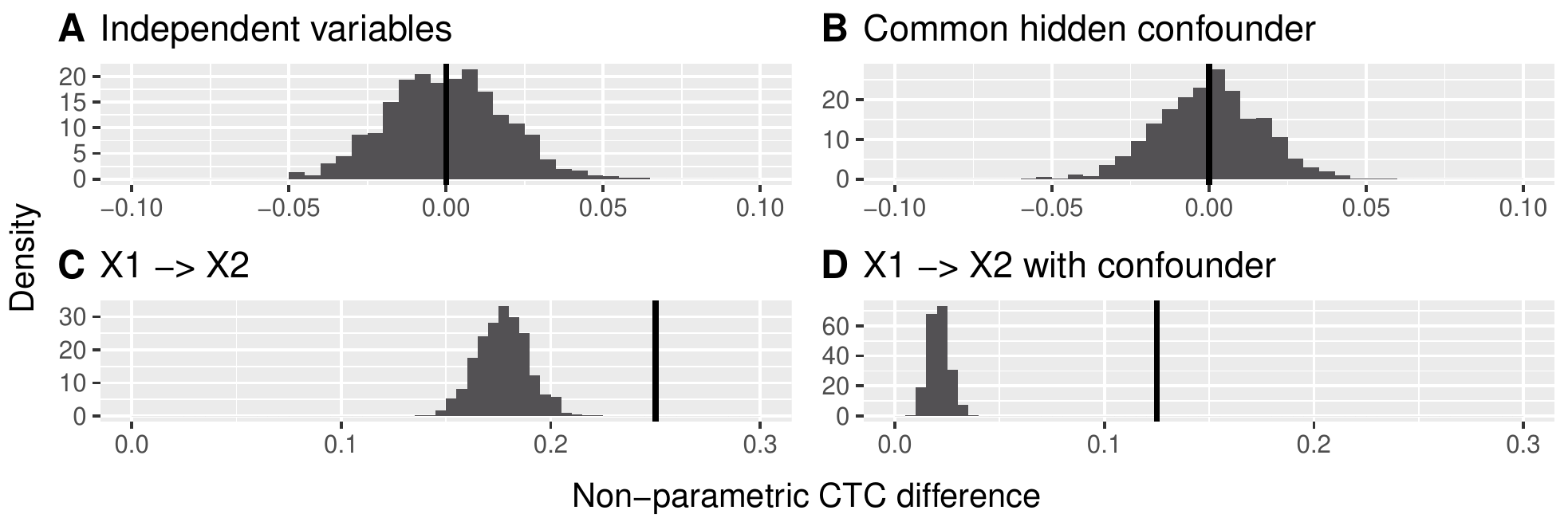}
\caption{Histogram of $\hat{\Delta}_{1,2}$ for $t_4$ distributed noise variables, for the four causal configurations. Lines indicate $\Delta_{1,2}=\Gamma_{1,2}-\Gamma_{2,1}$.}
\label{fs:dift4npctc}
\end{figure}

Figure~\ref{fs:npctc} shows the sample distributions of the non-parametric estimators $\hat{\Gamma}_{1,2}$ and $\hat{\Gamma}_{2,1}$ for all four causal structures, for the $t_4$, ${\rm Pareto}(1,2)$ and ${\rm LogN}(0,1)$ noise distributions, respectively. The true coefficient values $\Gamma_{1,2}$ and $\Gamma_{2,1}$ are obtained using~(2). %
Figure~\ref{fs:dift4npctc} shows the sample distribution of the coefficient difference estimator $\hat{\Delta}_{1,2}:=\hat{\Gamma}_{1,2} - \hat{\Gamma}_{2,1}$ for the $t_4$ noise distribution.

\subsection{LGPD Causal Tail Coefficient with Post-Fit and Constrained Fit Corrections}

\begin{figure}[p]
\centering
\includegraphics[width=0.93\textwidth]{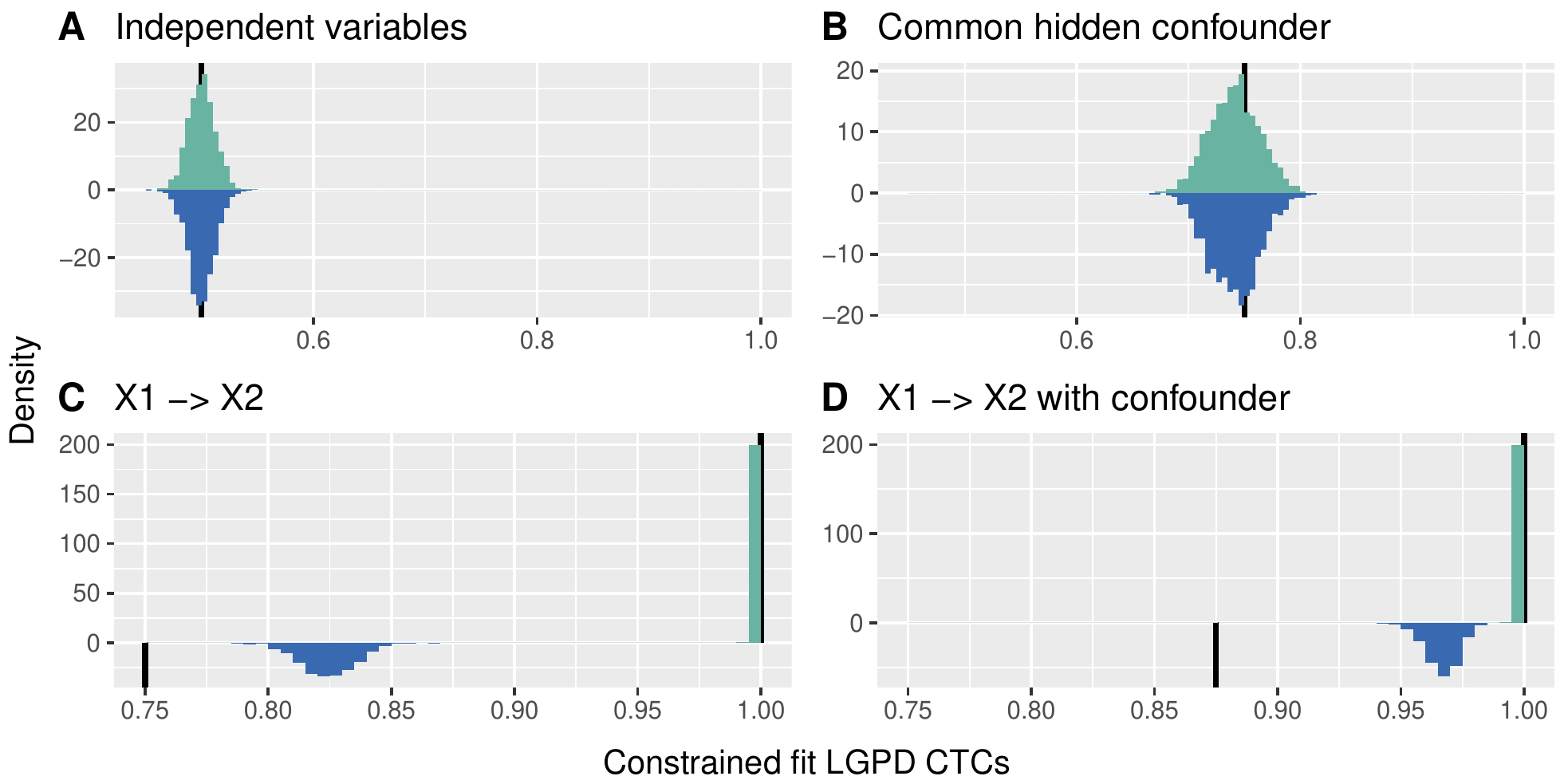}
\caption{Histograms of $\hat{\Gamma}_{1,2\mid H}^{\rm GPD}$ (turquoise) and $\hat{\Gamma}_{2,1\mid H}^{\rm GPD}$ (blue) with constrained fit for $t_4$ distributed noise variables, for the four causal configurations. Half-lines (black) indicate $\Gamma_{1,2}$ and $\Gamma_{2,1}$.}
\label{fs:t4cstrlgpdctc}
\end{figure}

\begin{figure}[p]
\centering
\includegraphics[width=0.93\textwidth]{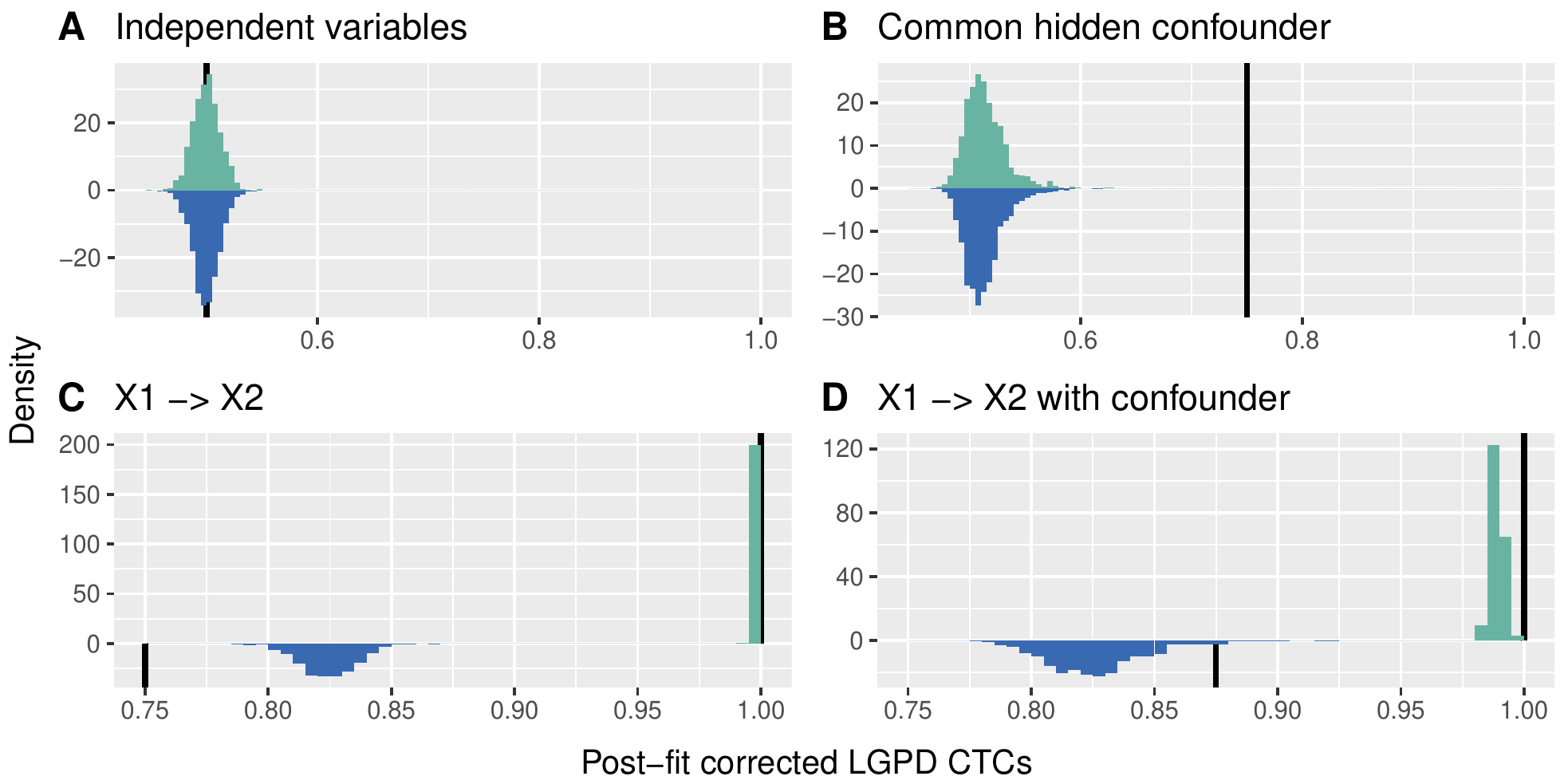}\\%
\includegraphics[width=0.93\textwidth]{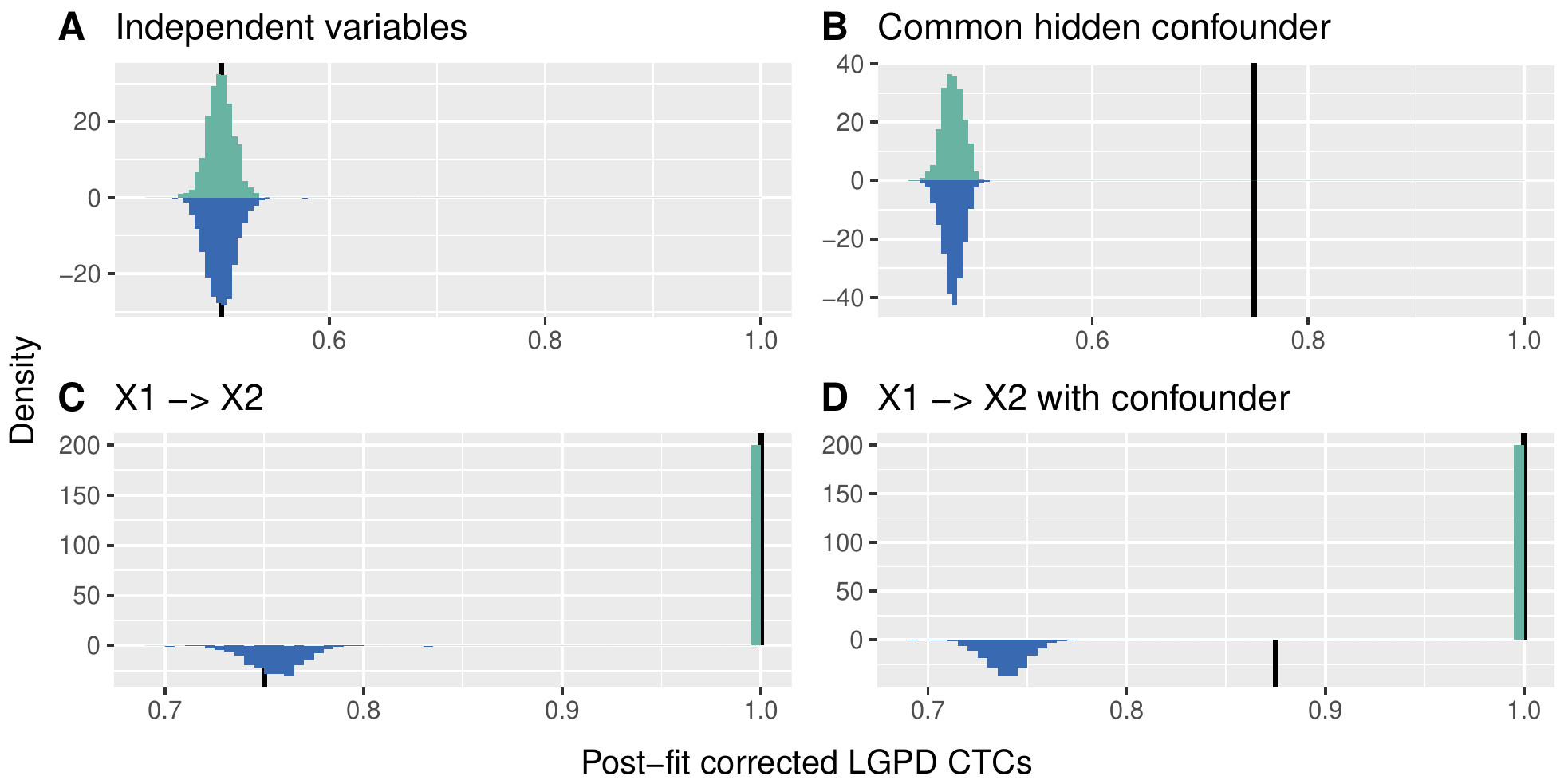}\\%
\includegraphics[width=0.93\textwidth]{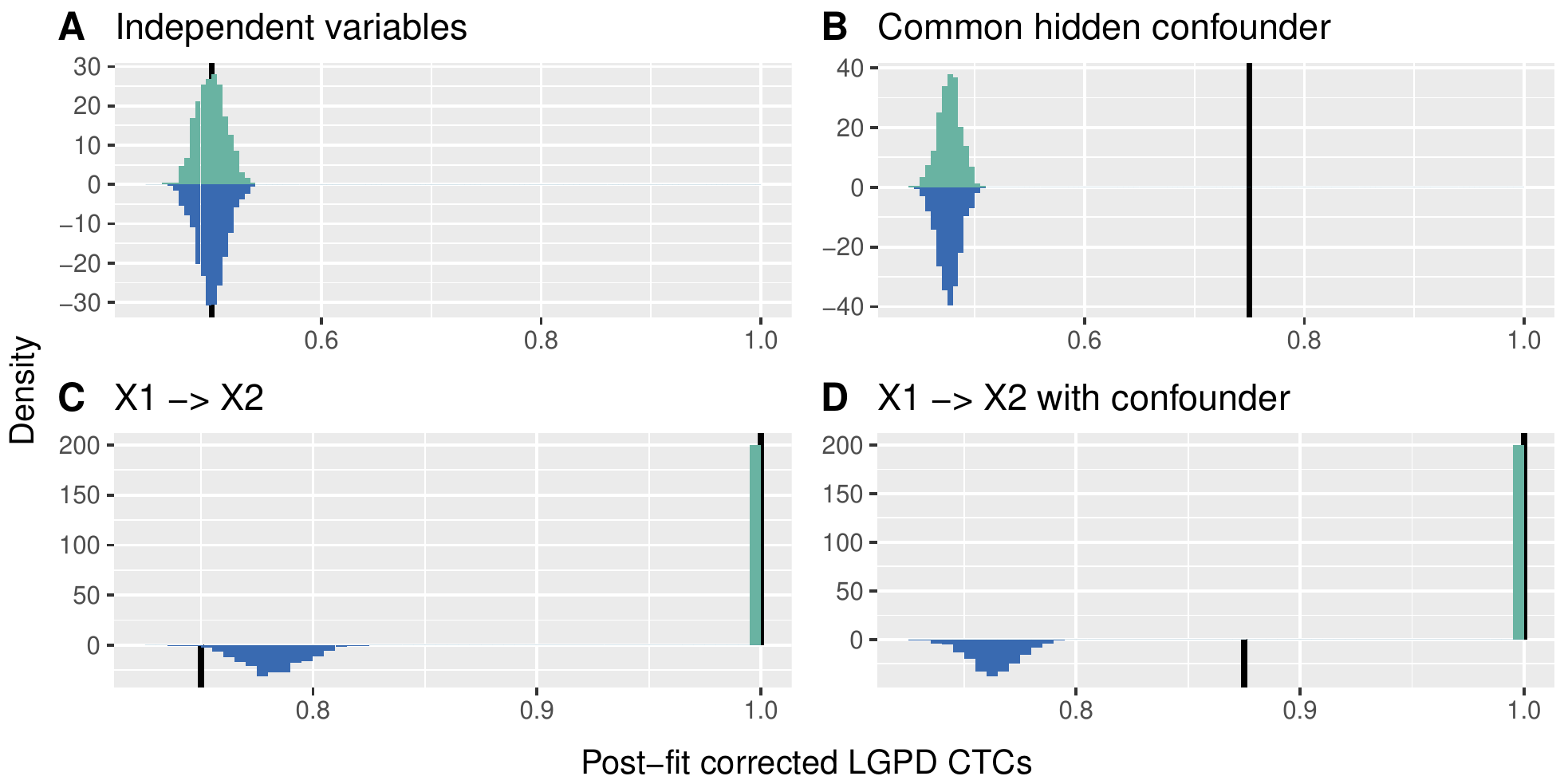}%
\caption{Histograms of $\hat{\Gamma}_{1,2\mid H}^{\rm GPD}$ (turquoise) and $\hat{\Gamma}_{2,1\mid H}^{\rm GPD}$ (blue) with post-fit correction for $t_4$ (top four panels), ${\rm Pareto}(1,2)$ (middle four panels) and ${\rm LogN}(0,1)$ (bottom four panels) distributed noise variables, for the four causal configurations. Half-lines (black) indicate $\Gamma_{1,2}$ and $\Gamma_{2,1}$.}
\label{fs:pflgpdctc}
\end{figure}

Figure~\ref{fs:t4cstrlgpdctc} shows the sample distribution of $\hat{\Gamma}_{1,2\mid H}^{\rm GPD}$ and $\hat{\Gamma}_{2,1\mid H}^{\rm GPD}$ with the constrained fit, for a comparable confounder tail. 
Figure~\ref{fs:pflgpdctc} shows the sample distribution of $\hat{\Gamma}_{1,2\mid H}^{\rm GPD}$ and $\hat{\Gamma}_{2,1\mid H}^{\rm GPD}$ with post-fit correction for all four causal configurations, for the $t_4$, ${\rm Pareto}(1,2)$ and ${\rm LogN}(0,1)$ noise distributions.

\section{Application Results for Competitors}\label{sm:applicompetitors}

Table~\ref{t:swisscompet} shows the  causal coefficients between the discharge station pairs estimated using ICA-LiNGAM, with and without considering the average catchment precipitation variable.

\begin{table}[p]
\centering
\caption{Linear causal coefficients for the discharge station pairs estimated with the ICA-LiNGAM algorithm using either the station pair only (LiNGAM, two variables) or the station pair and precipitation (LiNGAM-$H$, three variables). Non-null values indicate significant causal effects. The arrows indicate the estimated direct causal directions between the stations.\label{t:swisscompet}}
\nprounddigits{2}
\begin{tabular}{ll|n{1}{2}|n{1}{2}|}
Stations & Pair type & \text{LiNGAM}  & \text{LiNGAM-}$H$ \\
\hline
43-62 & causal    & 1.91929474292931$^\rightarrow$ & 2.02147993659115$^\rightarrow$ \\
42-63 & causal    & 2.07879186262232$^\rightarrow$ & 2.20875242829659$^\rightarrow$ \\
36-63 & causal    & 3.28936394744567$^\rightarrow$ & 3.60555863168095$^\rightarrow$ \\
24-61 & causal    & 2.96189249513038$^\rightarrow$ & 3.02797522258647$^\rightarrow$ \\
44-61 & causal    & 2.66344886172686$^\rightarrow$ & 2.83230029304895$^\rightarrow$ \\
22-38 & causal    & 2.35242595575385$^\rightarrow$ & 2.35242595575385$^\rightarrow$ \\
22-35 & causal    & 2.54841558325143$^\rightarrow$ & 2.54841558325143$^\rightarrow$ \\\hline
30-45 & non-caus. & 0.844373945195069$^\rightarrow$ & 0.87138638378129$^\rightarrow$ \\
36-39 & non-caus. & 0.6647466$^\leftarrow$ & 0.6563216$^\leftarrow$ \\
42-34 & non-caus. & 1.392225$^\leftarrow$ & 1.2881381$^\leftarrow$ \\
42-34$^*$ & non-caus. & 1.392225$^\leftarrow$ & 1.392225$^\leftarrow$ \\
32-33 & non-caus. & 0.593040513333397$^\rightarrow$ & 0.535594175755528$^\rightarrow$ \\
62-63 & non-caus. & 1.02472643450335$^\rightarrow$ & 1.04797361573715$^\rightarrow$ \\
57-60 & non-caus. & 0.68239595737648$^\rightarrow$ & 0.673070717007456$^\rightarrow$ \\
13-14 & non-caus. & 0.5025991$^\leftarrow$ & 1.10324474969347$^\rightarrow$ \\
17-22 & non-caus. & 1.79725395000552$^\rightarrow$ & 1.69241757070694$^\rightarrow$ \\
12-21 & non-caus. & 1.03944101513775$^\rightarrow$ & 1.07511701685805$^\rightarrow$ \\
26-28 & non-caus. & 0.7487394$^\leftarrow$ & 0.7193701$^\leftarrow$ \\
27-31 & non-caus. & 0.535030227721156$^\rightarrow$ & 0.660891637427842$^\rightarrow$ \\
23-39 & non-caus. & 0.254151192169057$^\rightarrow$ & 0.183805221682078$^\rightarrow$ \\
23-35 & non-caus. & 0.423726060357545$^\rightarrow$ & 0.356553057370821$^\rightarrow$ \\
\hline
\end{tabular}
\npnoround
\end{table}

\end{appendix}

\end{document}